\documentclass[authoryear,12pt]{elsarticle}
\usepackage{amsmath,amssymb,amsthm}
\usepackage[ruled]{algorithm2e}
\usepackage{graphicx}
\usepackage{color}
\usepackage{xcolor}
\usepackage{booktabs}
\usepackage{caption}
\usepackage{subcaption}
\usepackage{soul}
\usepackage{xurl}
\graphicspath{{fig/}}
\usepackage{algorithmic}

\usepackage{setspace}
\usepackage[T1]{fontenc}				
\usepackage[nomarkers]{endfloat}        
\usepackage{placeins}
\usepackage{ragged2e}

\usepackage{listings}
\usepackage{xcolor}
\definecolor{codegreen}{rgb}{0,0.6,0}
\definecolor{codegray}{rgb}{0.5,0.5,0.5}
\definecolor{codepurple}{rgb}{0.58,0,0.82}
\definecolor{backcolour}{rgb}{0.95,0.95,0.92}
\lstdefinestyle{mystyle}{
    backgroundcolor=\color{backcolour},   
    commentstyle=\color{codegreen},
    keywordstyle=\color{magenta},
    numberstyle=\tiny\color{codegray},
    stringstyle=\color{codepurple},
    basicstyle=\ttfamily\footnotesize,
    breakatwhitespace=false,         
    breaklines=true,                 
    captionpos=b,                    
    keepspaces=true,                 
    numbers=left,                    
    numbersep=5pt,                  
    showspaces=false,                
    showstringspaces=false,
    showtabs=false,                  
    tabsize=2
}
\allowdisplaybreaks

\newtheorem{remark}{Remark}

\journal{Elsevier}
\begin{document}
\baselineskip 24pt
\begin{frontmatter}
\author[1]{Zihao Wang}
\author[1,2]{Zhe Wu\corref{a}}
\cortext[a]{Corresponding author. E-mail: wuzhe@nus.edu.sg.\\Please refer to \url{https://github.com/killingbear999/chemical-reactor-foundation-model} for source codes.}

\address[1]{Department of Chemical and Biomolecular Engineering, National University of Singapore, 117585, Singapore}
\address[2]{Artificial Intelligence Institute, National University of Singapore, 117585, Singapore}
\title{Towards Foundation Model for Chemical Reactor Modeling: Meta-Learning with Physics-Informed Adaptation}

\begin{abstract}
Developing accurate models for chemical reactors is often challenging due to the complexity of reaction kinetics and process dynamics. Traditional approaches require retraining models for each new system, limiting generalizability and efficiency. In this work, we take a step toward foundation models for chemical reactor modeling by introducing a neural network framework that generalizes across diverse reactor types and rapidly adapts to new chemical processes. Our approach leverages meta-learning to pretrain the model on a broad set of reactor dynamics, enabling efficient adaptation to unseen reactions with minimal data. To further enhance generalizability, we incorporate physics-informed fine-tuning, ensuring physically consistent adaptation to new reactor conditions. Our framework is evaluated across three integer-order fundamental reactor types - continuous stirred tank reactors, batch reactors, and plug flow reactors - demonstrating superior few-shot adaptation compared to conventional data-driven, physics-informed, and transfer learning approaches. By combining meta-learning with physics-informed adaptation, this work lays the foundation for a generalizable modeling framework, advancing the development of foundation models for chemical engineering applications.
\end{abstract}
	
\end{frontmatter}

\section{Introduction}
\label{sec_introduction}

Research on foundation models has made significant progress in recent years. The term ``foundation model'' refers to any data-driven model trained on broad data, capable of adaptation (e.g., fine-tuning) to a broad spectrum of downstream tasks \citep{bommasani2021opportunities}. The success of foundation models in computer science, such as ChatGPT in natural language processing \citep{achiam2023gpt} and Sora in computer vision \citep{brooksvideo}, has spurred their applications in traditional engineering disciplines such as material science \citep{takeda2023foundation, miret2024llms} and biomolecular engineering (i.e., predict molecular properties using Simplified Molecular Input Line Entry System (SMILES) representation) \citep{ahmad2022chemberta, yang2024batgpt}, natural sciences such as astronomy \citep{lanusse2023astroclip}, physics \citep{mccabe2023multiple, herde2024poseidon} and biology \citep{bryant2022improved, hayes2024simulating}, and scientific machine learning \citep{hao2024dpot}.

However, research on foundation models in chemical engineering remains limited but is increasingly recognized as critical \citep{decardi2024generative}. While some progress has been made in developing foundation models for chemistry-related scientific questions \citep{horawalavithana2022foundation} and for solving chemical and biological tasks such as biomedical question answering and molecular synthesis \citep{livne2023nach0}, much of this work frames the problem through a linguistic lens, primarily constructing language models for domain-specific question answering. To the best of our knowledge, no research has yet addressed the development of foundation models for chemical reactor modeling. To bridge this gap, we take a step toward developing a foundation model for chemical reactor modeling, designed to generalize across a wide range of chemical processes in classical reactor systems. Our goal is to enable efficient and accurate modeling of nonlinear reaction dynamics while reducing the need for system-specific retraining.

Dynamic process modeling, particularly for chemical reactions in real-world reactors, plays a crucial role in applications such as process monitoring, optimization, and advanced process control \citep{wu2019machine1, hassanpour2020integrating, ellis2020encoder, jalanko2021adaptive, li2024machine}. Three idealized models are used to estimate the most important process variables of different chemical reactors: continuous stirred tank reactors (CSTRs), batch reactors (BRs), and plug flow reactors (PFRs). All three types of chemical reactors can be described using a set of ordinary differential equations (ODEs) or partial differential equations (PDEs), commonly referred to as first-principles models \citep{fogler2010essentials}. However, the parameters of these first-principles models vary depending on the specific reactions and reactors involved, and they can be difficult to determine accurately in real-world applications. Thus, there has been a transition to utilize data-driven methods, such as machine learning-based models, to capture the system dynamics of chemical reactions. These methods include least squares regression \citep{bhadriraju2019machine, tang2023data}, feedforward neural networks \citep{liu2024state}, recurrent neural networks (RNNs) \citep{wu2019machine1, wu2019machine2}, universal differential equation-based hybrid models \citep{bangi2022physics}, physics-informed RNNs (PIRNNs) \citep{zheng2023physics}, neural ordinary differential equations \citep{sorourifar2023physics}, and input convex neural networks \citep{wang2025real}. However, a significant hurdle in employing deep learning within scientific domains is the scarcity of experimental data, which is normally costly and time-consuming to acquire via physical experimentation (e.g., it is resource-intensive to collect data and build neural network models for every new chemical reaction in the reactor). To mitigate this challenge, transfer learning-based approaches have been suggested \citep{xiao2023modeling, xiao2024optimization}. However, for transfer learning-based methods, the process of identifying analogous reactions with adequate data, pre-training, and then transferring to the designated reaction must be repeated for each new reaction with a considerable number of samples, typically hundreds or thousands \citep{zhuang2020comprehensive}. 

\begin{figure}[ht]
\centering
\includegraphics[width=0.8\textwidth]{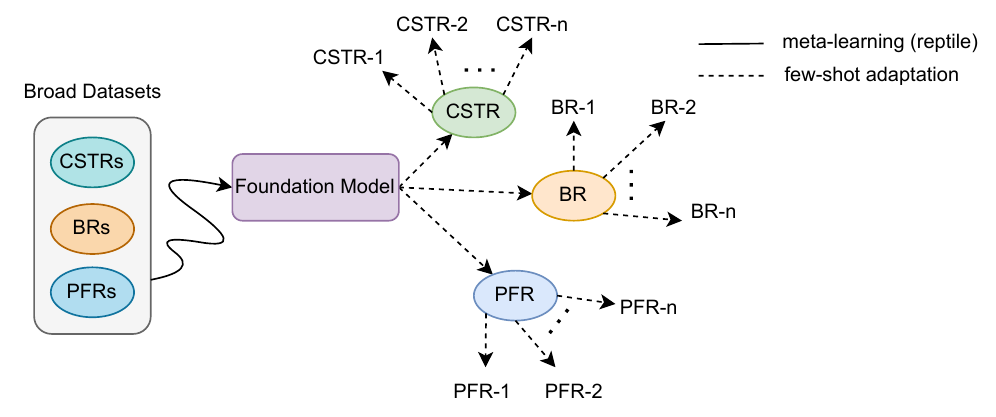}
\caption{High-level system architecture of ``foundation model'' for various chemical reactors.}
\label{fig_high_level}
\end{figure}

\begin{figure}[ht]
\centering
\includegraphics[width=\textwidth]{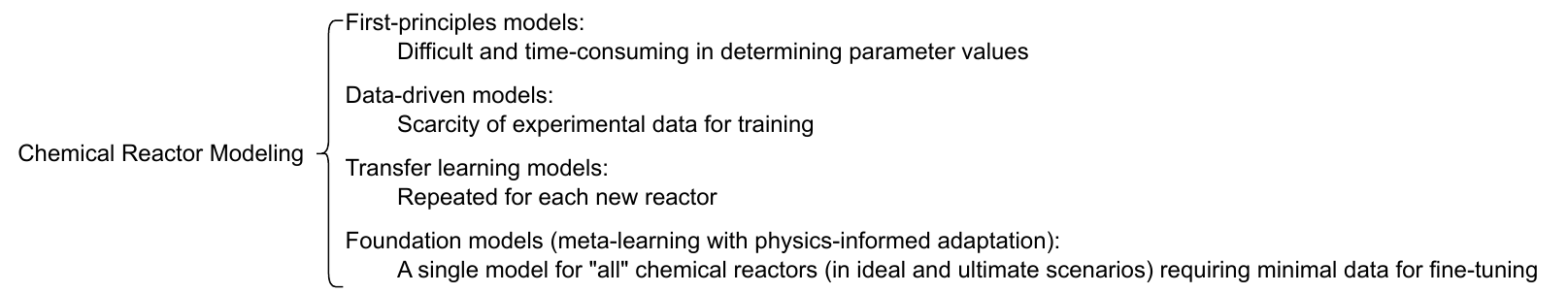}
\caption{Limitations of prior approaches and advantages of the proposed method.}
\label{fig_contribution}
\end{figure}

This raises a fundamental question: can we develop a foundation model that rapidly adapts to any new reaction for chemical reactor modeling? To take a step toward this goal, we propose a model trained on a diverse dataset of integer-order chemical reactions in CSTRs, BRs, and PFRs. Firstly, we employ a meta-learning technique to build the model, specifically Reptile \citep{nichol2018reptile}. After acquiring the model through Reptile, a conventional method to adapt the model to any new processes is via few-shot adaptation (the term ``shot'' refers to the number of data samples used to fine-tune the neural network). However, conventional data-driven adaptation in few-shot settings may result in poor generalization for nonlinear systems, due to the complexity of their state-space dynamics. To enhance the generalizability of the model and reduce the data required for adaptation, we draw inspiration from the success of physics-informed neural networks \citep{raissi2019physics}. Specifically, we adopt a physics-informed adaptation approach by integrating domain knowledge, such as governing equations (e.g., mass and energy balances), into the adaptation process. Simulation data of hundreds of different reactions in three classical reactors is used to develop the model that will be further adapted to new reaction processes with limited data available. It is demonstrated that the adaptation process from the model to a specific reaction is efficient, requiring only a few shots for successful adaptation. The high-level system architecture is shown in Fig. \ref{fig_high_level} and the limitations of prior approaches and advantages of the proposed method are summarized in Fig. \ref{fig_contribution}. In summary, the contributions of this work can be summarized in two main points:
\begin{itemize}
    \item We introduce, for the first time in the literature, the concept of a foundation model specifically for chemical reactor modeling. This work serves as a significant first step toward developing a comprehensive foundation model in this field.
    \item We present meta-learning and physics-informed adaptation as key enabling techniques, demonstrating their effectiveness in facilitating rapid adaptation to unseen chemical reactors within a pool of ideal basic single-step chemical reactors (i.e., CSTRs, BRs, and PFRs).
\end{itemize}

To avoid confusion and overstatement, we clarify that this work does not claim to fully develop a foundation model for chemical reactor modeling. Rather, it represents a step toward such models by proposing a framework in which meta-learning and physics-informed adaptation serve as key enablers. Accordingly, we use the term ``foundation model'' in quotation marks throughout the rest of this work to reflect this positioning. The rest of this article is structured as follows: Section \ref{sec_related} reviews the current works on physics-informed learning and meta-learning. Section \ref{sec_pi_reptile} presents the algorithms for the Reptile-based ``foundation model'' with physics-informed adaptation. Section \ref{sec_designs} outlines the formulation of CSTRs, BRs, and PFRs, covering reactor modeling and the design of physics-informed loss functions. Section \ref{sec_results} assesses the few-shot performance of the proposed method on unseen fixed-order CSTRs, BRs, and PFRs. Section \ref{sec_orders} extends the method to incorporate random integer reaction orders. Finally, Section \ref{sec_limitations} addresses the limitations of the current design and proposes directions for future research.

\begin{remark}
The effectiveness of transfer learning heavily relies on the similarity between the data distributions of the target and source tasks. In practice, finding a source reaction closely aligned with the target reaction is often challenging. A common approach to address this challenge involves conducting a few experiments under various conditions to gather data and develop a first-principles model through parameter optimization via regression. Once this model is established, it can be used to simulate a larger dataset, which then serves to pretrain a neural network for transfer learning, covering the entire operational range of the target reaction. However, transfer learning that depends on creating first-principles models to generate simulation datasets for each new reaction can be time-consuming, often taking hours or even days, and its effectiveness is highly contingent on the accuracy of the developed first-principles model. In contrast, our ``foundation model'' delivers high performance within seconds, \textbf{eliminating the need for dependencies on first-principles models}.
\end{remark}

\section{Meta-Learning and Physics-Informed Learning}
\label{sec_related}
Meta-learning algorithms, such as Model-Agnostic Meta-Learning (MAML) \citep{finn2017model} and its variants \citep{finn2018probabilistic, finn2019online, khodak2019adaptive, rajeswaran2019meta}, were initially developed for learning-to-learn tasks. In short, MAMLs are training algorithms designed to optimize neural network weights, facilitating easy adaptation to new tasks with just a few shots from those tasks \citep{vettoruzzo2024advances}. They have since been applied to provide initialization for physics-informed neural networks (PINNs) to accelerate solving PDEs \citep{liu2022novel, qin2022meta, penwarden2023metalearning}. Moreover, a neural operator and its variants \citep{li2020fourier, wang2021learning, li2024physics} have been proposed to learn operators mapping between infinite dimensional function spaces for solving parametric PDEs.

Our goal and approach differ from these works. Due to the nature of our tasks, we use an RNN as the base model to develop a foundational model for generic chemical reactors (i.e., different parameterized ODE systems). Unlike conventional PINNs, which involve multiple applications of automatic differentiation, we fine-tune the meta-learned RNN using a physics-informed loss term computed with numerical methods, such as the forward Euler method, which requires fewer computational resources. Moreover, unlike foundation models in natural language processing (NLP) and computer vision (CV) - such as ChatGPT \citep{achiam2023gpt} - which typically rely on self-supervised learning or reinforcement learning from human feedback (RLHF), our approach is grounded in meta-learning and physics-informed learning. These methods enable efficient adaptation to new chemical processes with limited data and are well-suited to the scientific modeling context. While our methodology differs from those used in large-scale language and vision models, it aligns with the core principles of foundation models: the ability to learn generalizable representations and adapt across tasks with minimal retraining. Although foundation models in NLP and CV are often evaluated across diverse task types, our model is designed to generalize across a wide range of reactor configurations and reaction systems within chemical process modeling. This eliminates the need to train separate models for each reactor type and supports broad adaptability within a relevant domain. Additionally, the integration of physics-informed adaptation ensures practical applicability in data-scarce settings, where zero-shot transfer may be unrealistic.

Furthermore, we believe that a successful foundation model should support both few-shot learning and continual learning, and existing literature indicates that meta-learning has significant potential for enabling continual learning \citep{riemer2018learning, javed2019meta, wang2024comprehensive}.

\begin{remark}
While many foundation models - such as those described by \citet{bommasani2021opportunities} - are large in scale, size alone does not define a foundation model. Its defining characteristics are generalization across diverse tasks, few-shot adaptability, and transferable representations. A central requirement is pretraining on a broad and diverse dataset, which we address by training on 1,500 simulated chemical systems encompassing CSTRs, BRs, and PFRs. Our model demonstrates scalability and extensibility by generalizing beyond its training distribution to unseen reaction kinetics, operating conditions, and reactor configurations. It is not limited to a single, narrowly defined task but adapts across multiple reactor systems without retraining. This aligns with emerging interpretations of foundation models in scoped, domain-relevant contexts. It should be noted that our model generalizes beyond specific training tasks to different reactors and conditions, consistent with the foundation model paradigm within a targeted domain. However, we do not claim generalization to a variety of process systems engineering tasks such as reactor design, fault detection, or control, as these tasks are fundamentally distinct due to task heterogeneity (e.g., classification vs. regression), differing mathematical structures (e.g., steady-state vs. dynamic models), and divergent modeling objectives.
Importantly, while foundation models are expected to support diverse tasks, transferability is meaningful only when new tasks are aligned with the pretraining objective. Therefore, robust generalization within chemical reactor modeling represents a necessary and tractable first step. 
\end{remark}

\begin{remark}
Rather than aiming to develop a universal foundation model for chemical engineering or process systems engineering, we position our model as a foundation model specifically for chemical reactor modeling, a coherent and practically significant subdomain.  In this context, the downstream tasks correspond to different types of chemical reactors, each with distinct dynamics and kinetics. This scoping aligns with the framing adopted in recent literature, such as the work by \cite{allen2024learning} on machine-learned interatomic potentials using meta-learning. In that study, the term ``foundation model'' is applied within the specific context of potential energy prediction, rather than across the entire domain of materials modeling. Similarly, we use the term in a domain-relevant manner to reflect our model's ability to generalize across diverse reactor systems without task-specific retraining. Thus, our work contributes to the emerging application of foundation models in scientific machine learning, demonstrating how such models can be tailored to specialized domains while maintaining key properties of generalization and transferability.
\end{remark}

\section{Methodology}
\label{sec_pi_reptile}
\subsection{Class of Systems}
In this work, we consider chemical processes that can be described by a specific class of nonlinear ODEs, including those derived from PDEs using methods such as the method of lines:
\begin{equation}\label{eq:classofsystems}
\dot{x} = F(x,u)
\end{equation}
where $x \in \mathbb{R}^n $ denotes the state vector (e.g., temperature, concentration, etc.), $u \in \mathbb{R}^m $ is the control input vector (e.g., adjusting feed rate and composition, etc.). $F : D \times U  \to \mathbb{R}^n $ is a continuously differentiable function, where $D \subset \mathbb{R}^n $ and $U \subset \mathbb{R}^m $ are compact and connected subsets that contain an open neighborhood of the origin, respectively. Since the explicit parameter values of first-principles models may not be available for complex real-world chemical reactors, our goal is to develop a neural network for the nonlinear system of Eq. \eqref{eq:classofsystems}.

\subsection{Reptile for Meta-Learning}
We begin by introducing the meta-learning algorithm used to build the ``foundation model''. Specifically, we employ Reptile \citep{nichol2018reptile}, a simplified version of MAML that prioritizes computational efficiency \citep{nichol2018first}. Unlike MAML, which relies on second-order derivatives, Reptile uses only first-order derivatives for meta-learning updates, making it more suitable for real-world applications. The primary goal of meta-training is to expose the model to a wide range of tasks, enabling it to learn initial parameters that facilitate rapid adaptation to new, unseen tasks. During this phase, Reptile optimizes the model's parameters for efficient adaptation to different tasks with minimal updates. In essence, Reptile guides the model toward a solution $\theta$ that closely approximates the manifold of optimal solutions for each task $\tau$, by minimizing the Euclidean distance through repeated task sampling, model updates, and parameter adjustments. Mathematically, Reptile identifies network parameters $\theta$ that minimize the Euclidean distance $\mathcal{D}(\theta, \mathcal{W}_\tau)$ for all tasks (i.e., locating a point near all solution manifolds), where $\mathcal{W}_\tau$ is the set of optimal parameters for task $\tau$:
\begin{equation}
    \min\limits_{\theta} \mathbb{E}_\tau[\frac{1}{2}\mathcal{D}(\theta, \mathcal{W}_\tau)^2]
\end{equation}
In our case, each task $\tau$ corresponds to a specific chemical process governed by the ODEs in Eq. \eqref{eq:classofsystems}.

At its core, Reptile focuses on learning how to quickly adapt to a variety of tasks, rather than specializing in any single one. Throughout meta-training, the model encounters diverse tasks and refines a parameter set optimized for rapid adaptation to new challenges. By simultaneously exposing the model to multiple tasks, Reptile identifies shared features across tasks, contrasting with conventional neural networks that focus on fitting individual tasks. This meta-learning process yields \textbf{adaptable parameters}, represented by $\theta$, which require minimal gradient updates to fine-tune for new tasks. Reptile excels at generalizing across tasks and quickly adapting to novel challenges, making it a robust framework for meta-learning. Conceptually, Reptile functions like a tug-of-war between task groups, pulling in different directions to find a balanced state.

\subsection{Reptile with Physics-Informed Adaptation}
Next, we develop a ``foundation model'' to learn the dynamics of various chemical reactions in three types of reactors. Our methodology comprises two distinct phases: meta-training and meta-testing. During meta-training (i.e., Algorithm \ref{alg_reptile}), we immerse the model (i.e., specifically, an RNN) in a diverse array of chemical reactions across various reactor types. Specifically, the meta-training process involves two steps. First, we obtain the optimal parameters $\mathcal{W}_\tau$ for a specific task $\tau$ using the mean squared error (MSE) and the \texttt{Adam} optimizer \citep{kingma2014adam}. Once $\mathcal{W}_\tau$ is determined, we update the network parameters $\theta$ using the Reptile update rule:
\begin{equation}
\theta \leftarrow \theta + \alpha(\mathcal{W}_\tau - \theta)
\end{equation}
where $\alpha$ is the linear schedule step size hyperparameter. This process is then repeated recursively for all tasks.

\begin{algorithm}[tb]
\caption{Reptile-Based ``foundation model'' Meta-Training}
\label{alg_reptile}
\textbf{Require}: $p(\mathcal{T}):$ distribution over tasks \\
\textbf{Require}: $\epsilon:$ meta-optimization step size hyperparameter \\
\textbf{Require}: $\alpha:$ linear schedule step size hyperparameter \\
\textbf{Require}: $x, y:$ training inputs and outputs respectively \\
\textbf{Require}: $M:$ a neural network
\begin{algorithmic}[1]
\STATE Initialize $\theta$.
\FOR {$iteration \leftarrow 1,2,\ldots,n_{tasks}$}
\STATE Sample one task $\tau \sim p(\mathcal{T})$.
\FOR {$innerepoch \leftarrow 1,2,\ldots,n_{epochs}$}
\STATE Let $\mathcal{W}_\tau \leftarrow \theta$.
\FOR {$i \leftarrow 1,2,\ldots,n_{batches}$}
\STATE $\widetilde{y_i} \leftarrow M(x_i)$. 
\STATE Compute loss $\mathcal{L}_{\tau_i}(\widetilde{y_i}, y_i)$ with respect to current batch.
\STATE Update $\widetilde{\theta} \leftarrow Adam(\mathcal{L}_{\tau_i}(\widetilde{y_i}, y_i), \mathcal{W}_\tau)$.
\ENDFOR
\STATE Compute $\alpha \leftarrow \epsilon (1 - \frac{iteration}{n_{tasks}})$.
\STATE Update $\theta \leftarrow \theta + \alpha (\mathcal{W}_\tau - \theta)$.
\ENDFOR
\ENDFOR
\end{algorithmic}
\end{algorithm}

In the subsequent meta-testing phase (i.e., Algorithm \ref{alg_pi_fewshots}), we select previously unseen chemical reactions and undertake physics-informed few-shot adaptations to them, respectively, leveraging the Reptile-based ``foundation model'' obtained during meta-training. It should be pointed out that we use few-shot data $x$ (i.e., $x \in \mathcal{X} \subset D$, where $\mathcal{X}$ denotes the set of few-shot training data) to compute the data-driven loss term $\mathcal{L}_{d}$ and use collocation points $x_c$ (i.e., $x_c \in \mathcal{X}_c \subset D$, where $\mathcal{X}_c$ denotes the set of collocation data points, and $\mathcal{X} \cap  \mathcal{X}_c= \emptyset$) to compute the physics-informed loss term $\mathcal{L}_{p}$ (a collocation point refers to a specific location within the domain of interest where the governing physics equations are enforced as part of the training process). As we do not utilize any labeled output data for $\mathcal{L}_{p}$, there is no need to perform physical experiments to gather additional data, which remains consistent with our few-shot setting. Moreover, it should be noted that we use \textbf{estimated} parameter values in $\mathcal{L}_{p}$ (i.e., fixed values) instead of their ground truth during adaptation (this differs from PIRNNs developed in our previous work \citep{zheng2023physics}, which use ground truth parameter values), as the explicit parameter values of the first-principles models are not available in real-world scenarios. It is important to note that using approximate parameter values in physics-based regularization can introduce a discrepancy between the learned model and the true system dynamics. However, it is important to recognize that if we had perfect knowledge of physical laws and system parameters, a purely data-driven approach would be unnecessary. Physics-informed machine learning is particularly valuable for systems where data is limited, and domain knowledge is incomplete or imprecise. One of the key benefits of physics-informed machine learning is its ability to balance data-driven learning with physical constraints, even when the exact parameter values are uncertain. Numerous studies have demonstrated that even when physics-based constraints are not entirely accurate, incorporating them can improve model generalization and robustness compared to purely data-driven methods trained on limited data. In this work, the benefits of using physics-informed machine learning include:
\begin{itemize}
    \item  Rather than enforcing strict adherence to potentially incorrect kinetics, the regularization term serves as a soft constraints that incorporates domain knowledge. The model primarily learns from data, with the regularization acting as an additional informative bias rather than an absolute constraint.
    \item In the absence of exact kinetic parameters, the chosen sampling intervals are based on literature values and physical intuition. Taking the center points ensures that the regularization remains unbiased and provides a reasonable parameter estimation, preventing systematic skewing toward extreme values.
    \item   Despite potential discrepancies, our results demonstrate that incorporating physics-based regularization improves the model’s generalizability and alignment with physical trends, validating its usefulness even with approximate parameter estimates.
\end{itemize}

The detailed formulation of physics-informed loss functions with respect to CSTRs, BRs, and PFRs will be presented in the following section, along with the parameterized first-principles models of CSTRs, BRs, and PFRs. 

\begin{algorithm}[tb]
\caption{Physics-Informed Few-shot Adaptation}
\label{alg_pi_fewshots}
\textbf{Require}: $p(\mathcal{T})$: distribution over tasks \\
\textbf{Require}: $M$: pre-trained ``foundation model'' \\
\textbf{Require}: $\gamma_1, \gamma_2$: loss function weighting hyperparameters \\
\textbf{Require}: $x, y$: few-shot inputs and outputs respectively \\
\textbf{Require}: $x_c$: collocation points
\begin{algorithmic}[1]
\STATE Sample an unseen task $\tau \sim p(\mathcal{T})$.
\STATE Load the parameters $\theta$ from $M$.
\FOR{$epoch \leftarrow 1,2,\ldots,n_{epochs}$}
\FOR{$i \leftarrow 1,2,\ldots,n_{batches}$}
\STATE $\widetilde{y_i} \leftarrow M(x_i)$. 
\STATE $\widetilde{y_{p}} \leftarrow M(x_c)$.
\STATE Compute loss $\mathcal{L}_{\tau_i} \leftarrow \gamma_1 \mathcal{L}_{d_{\tau}}(\widetilde{y_i}, y_i) + \gamma_2 \mathcal{L}_{p_{\tau}}(\widetilde{y_{p}}, x_c)$ with respect to current batch.
\STATE Update $\theta \leftarrow Adam(\mathcal{L}_{\tau_i}, \theta)$.
\ENDFOR
\ENDFOR
\end{algorithmic}
\end{algorithm}

\begin{remark}
We observed that employing the aforementioned approach to build a ``foundation model'' specifically for one type of reactor (e.g., CSTRs) can achieve a testing MSE on the order of $10^{-4}$ for normalized data easily with around 5 shots using a 2-layer RNN. However, based on our previous works \citep{wu2019machine1, wu2019machine2, wang2025real}, we observed that a testing MSE on \textbf{the order of $\mathbf{10^{-3}}$} for \textbf{normalized data} is indicative of satisfactory performance, rendering the model suitable for applications in process systems engineering such as neural network-based MPC.  
\end{remark}

\begin{remark}
While developing a ``foundation model'' for each type of reactor might yield better performance for that specific reactor, a unified ``foundation model'' for all types of reactors could offer a significant advantage in generalizability. Specifically, it can adapt to the same chemical reaction across different reactor types, which a single-type ``foundation model'' cannot achieve. This adaptability is crucial when the real-world system involves the same reaction but in a different reactor, as a single-type model lacks information about other reactor types. This motivates us to develop a unified ``foundation model'' that can handle all three types of chemical reactors, rather than creating separate models for each type (our goal is not to develop an extremely precise model).
\end{remark}

\begin{remark}
For applications requiring extremely low MSE on the order of $10^{-4}$ or $10^{-5}$, utilizing high-performing RNNs such as Long Short-Term Memory (LSTM) \citep{hochreiter1997long}, UniCORNN \citep{rusch2021unicornn}, and Linear Recurrent Units \citep{orvieto2023resurrecting} is advisable over traditional RNNs (despite the rise of Transformer-based approaches for time series forecasting, their effectiveness remains uncertain \citep{zeng2023transformers}). However, these advanced models come with a trade-off in computational efficiency. From an engineering standpoint, simplicity is crucial to ensure that the model can be easily integrated into existing industrial systems. Since simple RNNs can deliver satisfactory performance, this work aims to develop a network architecture based on simple RNNs that balances simplicity with adequate accuracy, rather than solely pursuing the optimal model architecture for accuracy. Additionally, increasing the number of collocation points can improve model performance, but significantly increases computational complexity, as indicated by the increase in the number of floating-point operations. In this work, we aim to maintain system simplicity to ensure computational efficiency, seeking a balance between computational efficiency and achieving a reasonable MSE for future real-time applications, such as neural network-based MPC.
\end{remark}

\section{Data Simulation for Chemical Reactors and Development of Physics-Informed Loss Term}
\label{sec_designs}
\begin{figure}[ht!]
\centering
\includegraphics[width=0.8\textwidth]{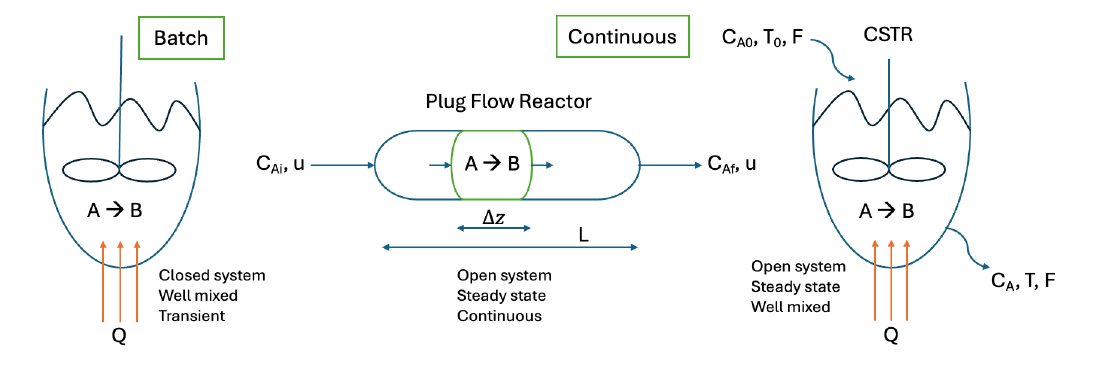}
\caption{Schematic of reactors.}
\label{fig_reactors}
\end{figure}

In this section, we present the formulation of CSTRs, BRs, and PFRs, alongside their corresponding physics-informed loss functions. In this work, we utilize the notations CSTRs, BRs, and PFRs to distinguish between different chemical reactions for easy reference (e.g., CSTRs denote multiple different CSTR-based chemical reactions). The schematic of the three reactors is shown in Fig. \ref{fig_reactors}. Furthermore, within the scope of this paper, we only considered elementary chemical reactions that transform reactant A to product B (irreversible reaction) for all three types of reactors. In this section and Section \ref{sec_results}, we consider CSTRs with a second-order reaction from A to B, BRs and PFRs with a first-order reaction from A to B to demonstrate the capability of the proposed modeling method for handling various elementary reactions with different orders and in different reactors. However, it should be mentioned that the proposed method can be generalized to more diverse reactions, provided that the dataset is available (we will demonstrate the adaptability of the proposed method in Section \ref{sec_orders} by extending it to accommodate different orders of reactions).

Regarding the chemistry of the system, our study does not focus on a specific chemical reaction but rather on a general class of reactions where species $A$ is converted into species $B$ (e.g., dimerization of Nitrogen Dioxide to Dinitrogen Tetroxide, dimerization of $\alpha$-Methylstyrene, dimerization of Isoprene, dimerization of Formaldehyde to Paraformaldehyde, etc.). This benchmark reaction is widely used in process modeling due to its nonlinear dynamics and the challenges associated with accurately characterizing reaction kinetics, particularly when precise kinetic parameters are unavailable. Our ``foundation model'' is designed to learn the behavior of a broad range of elementary reactions (integer-order) with kinetic parameters sampled within reasonable ranges. The objective is not to develop a model for a single, well-defined reaction but rather to create a generalizable framework adaptable to various chemical systems.

\subsection{CSTR Modeling and Physics-Informed Loss Term}
The CSTR is equipped with a heating jacket that supplies/removes heat at a rate $Q$. The second-order CSTR dynamic model is described by the following material and energy balance equations:
\begin{subequations}
\begin{align}
\frac{dC_A}{dt} = & \frac{F}{V}(C_{A0}-C_A)-k_0e^{\frac{-E_a}{RT}}C_A^2 \\
\frac{dT}{dt} = & \frac{F}{V}(T_0-T)+\frac{-\Delta H}{\rho_LC_p}k_0e^{\frac{-E_a}{RT}}C_A^2+\frac{Q}{\rho_LC_pV}
\end{align}  
\label{eq_cstr}
\end{subequations}
where $C_A$ is the concentration of reactant A, $V$ is the volume of the reacting liquid, $T$ is the temperature, $Q$ is the heat input rate, $C_{A0}$ is the inlet concentration of reactant A, $T_0$ is the inlet temperature, $F$ is the volumetric flow rate, $\rho_L$ is the constant density of the reacting liquid, $C_p$ is the heat capacity, $\Delta H$ is the enthalpy of reaction, $k_0$ is the pre-exponential constant, $E_a$ is the activation energy, and $R$ is the ideal gas constant. The manipulated inputs (i.e., control) in this system are represented by $\Delta C_{A0} = C_{A0} - C_{A0_s}$ and $\Delta Q = Q - Q_s$, such that the equilibrium point of the system is located at the origin of the state-space. The ranges of parameter values for the CSTRs are described in Table \ref{tab_cstr_values}.

\begin{remark}
In a chemical reactor (e.g., CSTR, PFR), an equilibrium point typically refers to a steady-state operating condition, where key state variables (e.g., concentrations, temperature) remain constant over time. In an ODE-based reactor model,
\begin{equation}
    \frac{dx}{dt} = f(x, u)
\end{equation}
where $x$ represents state variables (e.g., concentration, temperature) and $u$ represents inputs (e.g., flow rates, heating), the equilibrium point $x^*$ satisfies $f(x^*, u^*) = 0$, meaning the time derivatives vanish. This equilibrium point can be determined by solving the steady-state form of the reactor model (i.e., setting the governing equations to zero) or by numerically identifying fixed points of the system. A more detailed analysis can be found in our previous work \citep{wu2019machine1}.
\end{remark}

\begin{table}
\centering
\begin{tabular}{c|l}
\toprule[1pt]
\textbf{Parameter} & \textbf{Value Range} \\
\midrule
$F$ & $[0, 55]~m^3/hr$ \\
$V$ & $[0,11]~m^3$ \\
$Q_s$ & $0.0~kJ/hr$ \\
$T_0$ & $[0, 600]~K$ \\
$R$ & $8.314~kJ/kmol~K$ \\
$C_{A0_s}$ & $[0, 8]~kmol/m^3$ \\
$\rho_L$ & $[950, 1050]~kg/m^3$ \\
$C_p$ & $[0.219, 0.243]~kJ/kg~K$ \\
$E_a$ & $[4.75, 5.25] \times 10^4~kJ/kmol$ \\
$k_0$ & $[8.04, 8.88] \times 10^6~m^3/kmol~hr$ \\
$\Delta H$ & $[-1.21, -1.09] \times 10^4~kJ/kmol$ \\
\bottomrule[1pt]
\end{tabular}
\caption{Ranges of parameter values for CSTRs}
\label{tab_cstr_values}
\end{table}

In the context of CSTR modeling, the inputs consist of  {$[T_t, C_{A_t}, \Delta Q_t, \Delta C_{{A0}_t}]$} at the current time step $t$. The outputs entail the state trajectory $[T_{t+1, \ldots, n}, C_{A_{t+1, \ldots, n}}]$ over the subsequent $n$ time steps. Leveraging randomly drawn parameter values from a reasonable range and employing the material and energy balance equations depicted in Eq. \eqref{eq_cstr}, we perform numerical simulations of various CSTRs to collect their respective output sequences employing the explicit Euler method. The operational ranges of CSTRs and the parameters used for simulation are as follows: $T$, $C_{A}$, $\Delta Q$, $\Delta C_{A0}$ are evenly sampled from $[300, 600]~K$, $[0, 6]~kmol/m^3$, $[-5, 5] \times 10^5 ~kJ/hr$, and $[-3.5, 3.5]~kmol/m^3$ respectively. For the explicit Euler method, the integration time step $h_c = 1 \times 10^{-4}~hr$ is used to solve Eq. \eqref{eq_cstr}, and the manipulated inputs $\Delta Q$, $\Delta C_{A0}$ are changed every sampling period of $5 \times 10^{-3}~hr$. However, it is noted that the ranges of variables can be chosen appropriately for the ``foundation model'' depending on the operating conditions of interest.

\begin{remark}
We have tested different numerical integration methods, including Runge-Kutta and implicit Euler, and found that the differences in results are minimal. This is particularly true for the CSTR system, which we previously analyzed and confirmed is not significantly stiff \citep{wu2019machine1}. Additionally, we ensured that the chosen time step is sufficiently small, preventing numerical oscillations or divergence. Thus, the explicit Euler method provides reliable training data for our case.
\end{remark}

Subsequently, when implementing the ``foundation model'' to a new process (e.g., a new reaction in a CSTR), we will apply physics-informed adaptation method. Specifically, the physics-informed loss function for learning a new CSTR is the weighted sum of the data-driven loss term $\mathcal{L}_{d}$ and the physics-informed loss term $\mathcal{L}_{p}$. $\mathcal{L}_{d}$ takes the form of MSE of the actual output $y$ and the predicted output $\Tilde{y}$ using the actual few-shot data, as follows:
\begin{equation}
\label{eq_cstr_dd_loss}
    \mathcal{L}_{d} = \frac{1}{n}\sum_{i=1}^{n}(\Tilde{y_i} - y_i)^2
\end{equation}
where $y$ is the actual output and $\Tilde{y}$ is the predicted output.
$\mathcal{L}_{p}$ is defined as the weighted sum of two terms, namely the loss term with respect to predicted concentration $\mathcal{L}_{C_A}$ and the loss term with respect to predicted temperature $\mathcal{L}_{T}$,  using collocation points only. It is expressed as follows:
\begin{equation}
\label{eq_cstr_pi_loss}
    \mathcal{L}_{p} = \gamma_1 \mathcal{L}_{C_A} + \gamma_2 \mathcal{L}_{T} = \gamma_1 \frac{1}{n} \sum_{i=1}^{n}\mathcal{L}_{C_{A_{t+i}}}^2 + \gamma_2 \frac{1}{n} \sum_{i=1}^{n}\mathcal{L}_{T_{t+i}}^2
\end{equation}
where
\begin{subequations}
\label{eq_cstr_pi_loss_ca_t}
\begin{align}
\mathcal{L}_{C_{A_{t+i}}} = & \frac{d \widetilde{C_A}_{t+i}}{dt} - \frac{F}{V}(C_{A0_{t}} - \widetilde{C_{A}}_{t+i}) + k_0e^{\frac{-E_a}{R\widetilde{T}_{t+i}}}\widetilde{C_{A}}_{t+i}^2 \\
\mathcal{L}_{T_{t+i}} = & \frac{d \widetilde{T}_{t+i}}{dt} - \frac{F}{V}(T_0-\widetilde{T}_{t+i}) + \frac{\Delta H}{\rho_LC_p}k_0e^{\frac{-E_a}{R\widetilde{T}_{t+i}}}\widetilde{C_{A}}_{t+i}^2 - \frac{Q_{t}}{\rho_LC_pV}
\end{align}
\end{subequations}

Here, $\widetilde{C_A}$ and $\widetilde{T}$ represent the predicted outputs corresponding to the collocation points as inputs and we use the explicit Euler method to solve the ODEs. When operating a real-world CSTR, certain parameters such as $F$, $V$, $T_0$, $C_{A0_s}$, and $Q_s$ are typically known through expert knowledge and direct physical measurements. However, other critical parameters, such as $k_0$, $E_a$, $\Delta H$, $C_p$, and $\rho_L$, may not be directly measurable and therefore remain unknown. To compute $\mathcal{L}_{p}$, we utilize the actual values of $F$, $V$, $T_0$, $C_{A0_s}$, and $Q_s$, while \textbf{estimating} the values of $k_0$, $E_a$, $\Delta H$, $C_p$, and $\rho_L$ in Eq. \eqref{eq_cstr_pi_loss_ca_t}. Specifically, we use the median values of $k_0$, $E_a$, $\Delta H$, $C_p$, and $\rho_L$ from their respective sampling ranges.

\subsection{BR Modeling and Physics-Informed Loss Term}
The BR is equipped with a heating jacket that supplies/removes heat at a rate $Q$. The first-order BR dynamic model is described by the following material and energy balance equations:
\begin{subequations}
\begin{align}
\frac{dC_A}{dt} = & -k_0e^{\frac{-E_a}{RT}}C_A \\
\frac{dT}{dt} = & \frac{-\Delta H}{\rho_LC_p}k_0e^{\frac{-E_a}{RT}}C_A + \frac{Q}{\rho_LC_pV}
\end{align}
\label{eq_br}
\end{subequations}
where the notations follow those for CSTRs, and are omitted here. The manipulated inputs (i.e., control) in this system are represented by $\Delta Q = Q - Q_s$, such that the equilibrium point of the system is located at the origin of the state-space. The ranges of parameter values for   BRs are described in Table \ref{tab_br_values}. 

\begin{table}
\centering
\begin{tabular}{c|l}
\toprule[1pt]
\textbf{Parameter} & \textbf{Value Range} \\
\midrule
$V$ & $[0,11]~m^3$ \\
$R$ & $8.314~kJ/kmol~K$ \\
$C_p$ & $[0.219, 0.243]~kJ/kg~K$ \\
$E_a$ & $[4.75, 5.25] \times 10^4~kJ/kmol$\\
$\rho_L$ & $[950, 1050]~kg/m^3$ \\
$k_0$ & $[0, 9.31] \times 10^8~m^3/kmol~hr$ \\
$\Delta H$ & $[-1.21, -1.09] \times 10^4~kJ/kmol$ \\
\bottomrule[1pt]
\end{tabular}
\caption{Ranges of parameter values for BRs}
\label{tab_br_values}
\end{table}

In the context of BR modeling, to maintain uniform input dimensions akin to CSTRs, we implement padding on the inputs for BRs. Consequently, the final inputs comprise  {$[T_t, C_{A_t}, \Delta Q_t, \Delta Q_t]$}\footnote{It is important to highlight that the variable $\Delta Q_t$ is not repeated; rather, it appears due to padding along the feature dimension to ensure uniform input size for the model. This padding is necessary for processing sequences in the RNN while maintaining a consistent feature representation across different samples.} at the current time step $t$, while the outputs encapsulate the state trajectory $[T_{t+1, \ldots, n}, C_{A_{t+1, \ldots, n}}]$ over the subsequent $n$ time steps. Leveraging randomly drawn parameter values from a reasonable range and employing the material and energy balance equations depicted in Eq. \eqref{eq_br}, we conduct numerical simulations of various BRs to collect their respective output sequences employing the explicit Euler method. Specifically, the operational ranges of BRs as well as the parameters used for data simulation are as follows: $T$, $C_{A}$, $\Delta Q$ are evenly sampled from $[300, 600]~K$, $[0, 6]~kmol/m^3$, and $[-5, 5] \times 10^5 ~kJ/hr$, respectively. For explicit Euler method, the integration time step $h_c = 1 \times 10^{-4}~hr$ is used to solve Eq. \eqref{eq_br}, and the manipulated input $\Delta Q$ is changed every sampling period of $5 \times 10^{-2}~hr$.

The BR physics-informed loss function is the weighted sum of the data-driven loss term $\mathcal{L}_{d}$ and the physics-informed loss term $\mathcal{L}_{p}$. $\mathcal{L}_{d}$ and $\mathcal{L}_{p}$ take forms similar to those of CSTR (that is, Eq. \eqref{eq_cstr_dd_loss} and Eq. \eqref{eq_cstr_pi_loss}) but with different $\mathcal{L}_{C_{A_{t+i}}}$ and $\mathcal{L}_{T_{t+i}}$, where 
\begin{subequations}
\label{eq_br_pi_loss_ca_t}
\begin{align}
\mathcal{L}_{C_{A_{t+i}}} = & \frac{d \widetilde{C_A}_{t+i}}{dt} + k_0e^{\frac{-E_a}{R\widetilde{T}_{t+i}}}\widetilde{C_{A}}_{t+i} \\
\mathcal{L}_{T_{t+i}} = & \frac{d \widetilde{T}_{t+i}}{dt} + \frac{\Delta H}{\rho_LC_p}k_0e^{\frac{-E_a}{R\widetilde{T}_{t+i}}}\widetilde{C_{A}}_{t+i} - \frac{Q_{t}}{\rho_LC_pV}    
\end{align}
\end{subequations}

Given a new BR, we also assume that the actual values of $V$ and $Q$ are known, while the values of $k_0$, $E_a$, $\Delta H$, $C_p$, and $\rho_L$ remain unknown and will be \textbf{estimated} (i.e., using the median values of their respective sampling ranges) in Eq. \eqref{eq_br_pi_loss_ca_t} for model adaptation.

\subsection{PFR Modeling and Physics-Informed Loss Term}
The first-order PFR dynamic model is described by the following material and energy balance equations:
\begin{subequations}
\begin{align}
\frac{\partial C_A}{\partial t} = & -u\frac{\partial C_A}{\partial z} - k_0e^{\frac{-E_a}{RT}}C_A \\
\frac{\partial T}{\partial t} = & -u\frac{\partial T}{\partial z} + \frac{-\Delta H}{\rho_LC_p}k_0e^{\frac{-E_a}{RT}}C_A + \frac{U}{\rho_LC_pA}A_c(T_c-T) 
\end{align}
\label{eq_pfr}
\end{subequations}
To solve PDEs for the PFR model, we employ the method of lines. Specifically, we discretize the reactor in length to approximate the spatial derivatives through finite differences, which results in a set of coupled ODEs (i.e., discretizing reactor length $L$ into $N$ sub-parts). These ODEs are subsequently solved using explicit Euler method to obtain future states for one sampling period along the reactor length, as depicted in Eq. \eqref{eq_pfr_method}:
\begin{subequations}
\label{eq_pfr_method}
\begin{align}
\frac{dC_{A_N}}{dt} = & -u\frac{C_{A_N} - C_{A_{N-1}}}{Z_{N-1} - Z_N} - k_0e^{\frac{-E_a}{RT}}C_A \\
\frac{dT_N}{dt} = & -u\frac{T_N - T_{N-1}}{Z_{N-1} - Z_N} + \frac{-\Delta H}{\rho_LC_p}k_0e^{\frac{-E_a}{RT}}C_A + \frac{U}{\rho_LC_pA}A_c(T_c-T)   
\end{align}
\end{subequations}
where $Z$ is the discretized elements (i.e., $N-1$ evenly spaced numbers over a specified reactor length $L$), $U$ is the overall heat transfer coefficient, $u$ is the superficial velocity, $T_c$ is the cooling liquid temperature, $A$ is the surface area, $A_c$ is the cross-sectional area of the tubular reactor (i.e., this parameter is typically denoted as $A_t$ in chemical engineering, but we opt for the designation $A_c$ to prevent confusion with the subscript $t$ representing time steps). The notations for other parameters follows those in CSTRs and are omitted here. The manipulated input (i.e., control) in this system is represented by $\Delta T_c = T_c - T_{c_s}$. The ranges of parameter values for PFRs are described in Table \ref{tab_pfr_values}. 

\begin{table}
\centering
\begin{tabular}{c|l}
\toprule[1pt]
\textbf{Parameter} & \textbf{Value Range} \\
\midrule
$A$ & $[0, 0.022]~m^2$ \\
$A_c$ & $[0, 0.11]~m^2$ \\
$L$ & $1~m$ \\
$u$ & $[0, 4]~m/min$ \\
$U$ & $[0, 50]~kcal/m^2~K$ \\
$N$ & $10$ \\
$\rho_L$ & $[950, 1050]~kg/m^3$ \\
$R$ & $8.314~kJ/kmol~K$ \\
$T_{c_s}$ & $[273, 586]~K$ \\
$C_p$ & $[0.219, 0.243]~kJ/kg~K$ \\
$E_a$ & $[4.75, 5.25] \times 10^4~kJ/kmol$\\
$k_0$ & $[8.04, 8.88] \times 10^6~m^3/kmol~hr$ \\
$\Delta H$ & $[-1.21, -1.09] \times 10^4~kJ/kmol$\\
\bottomrule[1pt]
\end{tabular}
\caption{Ranges of parameter values for  PFRs}
\label{tab_pfr_values}
\end{table}

In the context of PFR modeling, to maintain uniform input dimensions akin to CSTRs and BRs, we implement padding on the inputs for PFRs. Furthermore, we exclusively monitor the output from a single discretized location over the sampling period. Specifically, in this study, we focus on tracking the output from the first discretized point, where $N = 1$. Consequently, the final inputs comprise  {$[T_t, C_{A_t}, \Delta T_{c_t}, \Delta T_{c_t}]$}\footnote{It is important to highlight that the variable $\Delta T_{c_t}$ is not repeated; rather, it appears due to padding along the feature dimension to ensure uniform input size for the model.} at the current time step $t$, while the outputs encapsulate the state trajectory $[T_{t+1, \ldots, n}, C_{A_{t+1, \ldots, n}}]$ over the subsequent $n$ time steps. Leveraging parameter values randomly drawn from a reasonable range and employing the material and energy balance equations depicted in Eq. \eqref{eq_pfr_method}, we perform numerical simulations of various PFRs to collect their respective output sequences, which involve employing the explicit Euler method. Specifically, the operational ranges of PFRs as well as the parameters used for data simulation are as follows: $T$, $C_{A}$, $\Delta T_c$ are evenly sampled from $[300, 500]~K$, $[0.5, 3]~kmol/m^3$, and $[100, 300]~K$, respectively. For explicit Euler method, the integration time step $h_c = 0.01~hr$ is used to solve Eq. \eqref{eq_pfr_method}, and the manipulated input $\Delta T_c$ is changed every sampling period of $0.1~hr$.

The PFR physics-informed loss function is the weighted sum of the data-driven loss term $\mathcal{L}_{d}$ and the physics-informed loss term $\mathcal{L}_{p}$. $\mathcal{L}_{d}$ and $\mathcal{L}_{p}$ take similar forms as those for CSTRs (i.e., Eq. \eqref{eq_cstr_pi_loss}) but with different $\mathcal{L}_{C_{A_{t+i}}}$ and $\mathcal{L}_{T_{t+i}}$, where
\begin{subequations}
\label{eq_pfr_pi_loss_ca_t}
\begin{align}
\mathcal{L}_{C_{A_{t+i}}} = & \frac{d \widetilde{C_A}_{t+i}}{dt} + u\frac{\widetilde{C_{A}}_{t+i} - C_{A_{t}}}{Z_1 - Z_0} + k_0e^{\frac{-E_a}{R\widetilde{T}_{t+i}}}\widetilde{C_{A}}_{t+i} \\ 
\mathcal{L}_{T_{t+i}} = & \frac{d \widetilde{T}_{t+i}}{dt} + u\frac{\widetilde{T}_{t+i} - T_{t}}{Z_1 - Z_0} - \frac{-\Delta H}{\rho_LC_p}k_0e^{\frac{-E_a}{R\widetilde{T}_{t+i}}}\widetilde{C_{A}}_{t+i} - \frac{U}{\rho_LC_pA}A_c(T_{c_{t}}-\widetilde{T}_{t+i})    
\end{align}
\end{subequations}

The calculation of the physics-informed loss term follows the method for CSTRs and BRs. Specifically, we utilize the actual values of $L$, $N$, $T_{c_s}$, $A$, $A_c$, $u$, and $U$, and \textbf{estimate} the values of $k_0$, $E_a$, $\Delta H$, $C_p$, and $\rho_L$ (i.e., using the median values of $k_0$, $E_a$, $\Delta H$, $C_p$, and $\rho_L$ from their respective sampling ranges) in Eq. \eqref{eq_pfr_pi_loss_ca_t} for model adaptation. Note that the terms $\frac{d \widetilde{C_A}_{t+i}}{dt}$ and $\frac{d \widetilde{T}_{t+i}}{dt}$ in $\mathcal{L}_{p}$ of CSTRs, BRs, and PFRs, can be computed using numerical methods such as the explicit Euler method. 


\begin{remark}
We apply padding to the inputs of BRs and PFRs to ensure a consistent input dimension for the ``foundation model'', accommodating all three types of reactors. Here, padding is not intended to improve results but rather to ensure uniform input dimensions, which is crucial for processing data in an RNN. Specifically, we pad along the feature dimension to maintain consistency across inputs. Without this, we would need to train separate RNNs for different input types, which contradicts our goal of developing a single unified model capable of handling various chemical reactions across different reactor types. Ablation studies revealed that padding with manipulated inputs (e.g., $\Delta Q$ for BRs and $\Delta T_c$ for PFRs) improves model performance compared to conventional padding methods, such as using natural values, e.g., $-1$, $0$, or $1$.
\end{remark}

\begin{remark}
Training PINNs can be notably challenging, and their performance is highly sensitive to the hyperparameters $\gamma_i$ governing the weighting of the loss terms. The weighting of the loss components was chosen to ensure that all loss terms remain within the same order of magnitude - a common practice in multi-objective optimization. Specifically, we assign a weight of $\gamma_1 = 1 \times 10^3$ to the data-driven loss $\mathcal{L}_{d}$,  reflecting our greater reliance on it as the primary learning signal. In contrast, the physics-informed losses $\mathcal{L}_{C_A}$ and $\mathcal{L}_{T}$ are assigned smaller weights, $\gamma_2 = 1 \times 10^{-2}$, and $\gamma_3 = 1 \times 10^{-5}$, respectively. This design reflects our intention for the model to primarily learn from data, with physics-based constraints serving as informative regularization rather than dominating the training process. These weights were selected based on empirical tuning through grid search over a predefined range (i.e., $\gamma_i \in [1 \times 10^{-5}, 1 \times 10^{5}])$, selecting the combination that yielded the lowest validation error. In summary, the total loss for CSTR, BR, and PFR models is formulated as:
\begin{equation}
\label{eq_cstr_loss}
\mathcal{L} = \gamma_1 \mathcal{L}_{d} + \gamma_2 \mathcal{L}_{C_A} + \gamma_3 \mathcal{L}_{T}
\end{equation}
The chosen weights gave the best performance across the three reactor types within the explored hyperparameter space, although they may not represent a global optimum.
\end{remark}

\begin{remark}
It is essential to highlight that we enforce constraints on the simulated output trajectories for all CSTRs, BRs, and PFRs to ensure that they remain within a physically plausible range. Specifically, we applied checks during data generation to detect and remove trajectories containing unrealistic values such as Inf (infinity), NaN (not-a-number), negative concentrations or temperatures (where physically infeasible), extremely large values, or values close to zero where nonzero quantities are expected. If a generated trajectory violated any of these conditions, we excluded the corresponding example and reran the simulation using a new set of randomly selected parameters within the ranges specified in Tables \ref{tab_cstr_values}, \ref{tab_br_values}, and \ref{tab_pfr_values}. Importantly, the numerical instabilities we refer to are not due to the use of the explicit Euler method, but instead stem from specific parameter combinations in the simulated chemical systems. For example, when the temperature is extremely low, the Arrhenius expression for the reaction rate can underflow, causing the reaction rate term to approach zero. This can result in degenerate system dynamics that are unrealistic or non-informative. The filtering process was therefore essential not only for removing unstable simulations but also for ensuring that the model is trained on physically meaningful and relevant trajectories. This contributes to the overall credibility and robustness of the foundation model development.
\end{remark}

\begin{remark}
The integration time step and the sampling period vary across CSTRs, BRs, and PFRs in this simulation study due to different kinetics and process dynamics. However, it is important to note that the ``foundation model'' can accommodate differing integration time steps and sampling periods for real-world reactors, provided the input dimensions remain consistent. This is due to the fact that the inputs to the neural network are merely numerical values, lacking any inherent physical interpretation for the network. Our approach does not require explicit architectural modifications because the RNN processes sequences of numerical values without relying on the physical interpretation of time steps. As long as the input dimension (number of features per timestep) and sequence length (maximum sequence length in a batch) remain consistent, the RNN can learn to map different reaction trajectories, regardless of the integration time step used during data generation. Specifically, The RNN is not inherently tied to a specific time discretization; it learns patterns based on numerical sequences rather than absolute time scales. Moreover, if different integration time steps result in varying trajectory lengths, padding or resampling would be necessary to ensure uniform input size. However, within a consistent input structure, the RNN can adapt to reactions evolving at different rates. However, while the RNN can implicitly learn representations across different time-step sizes, extreme differences in time scales (e.g., very stiff reactions) may still require additional considerations, such as adaptive sampling or time-step normalization.
\end{remark}

\section{Few-Shot Performance to Fixed-Order Unseen Reactions}
\label{sec_results}

We initiate the meta-training phase by training the Reptile-based ``foundation model'' on 500 CSTRs, 500 BRs, and 500 PFRs with different parameters representing various reactions in different settings of reactors (e.g., reactor size, flow rate, etc.). In actual deployment to industrial engineering processes, the ``foundation model'' is trained offline, and the costs associated with the meta-training process, such as dataset generation, are not a primary concern. Following this, in the meta-testing phase, we evaluate the few-shot performance of the pre-trained ``foundation model'' on 5 previously unseen CSTRs, 5 previously unseen BRs, and 5 previously unseen PFRs, respectively, using MSE as the evaluation metric. Specifically, during meta-testing, we fine-tune the ``foundation model'' using a limited number of samples\footnote{Each sample corresponds to a concentration-temperature pair at a single point in time, serving as an initial condition (i.e., the current state), while the RNN predicts the trajectory of future states.} from the designated reaction and assess its performance by computing the MSE over all samples from the same reaction, where the samples used for few-shot fine-tuning and the collocation points are randomly selected from the whole operational region of the designated reaction. In Figs. \ref{fig_cstr}, \ref{fig_br}, \ref{fig_pfr}, and \ref{fig_performance_orders}, the solid line represents the mean of 5 trials on different unseen reactions, while the shaded region indicates the standard deviation. The $y$-axis is presented on a \textbf{logarithmic scale} for clarity.

\begin{figure}[ht]
\centering
\includegraphics[width=0.8\textwidth]{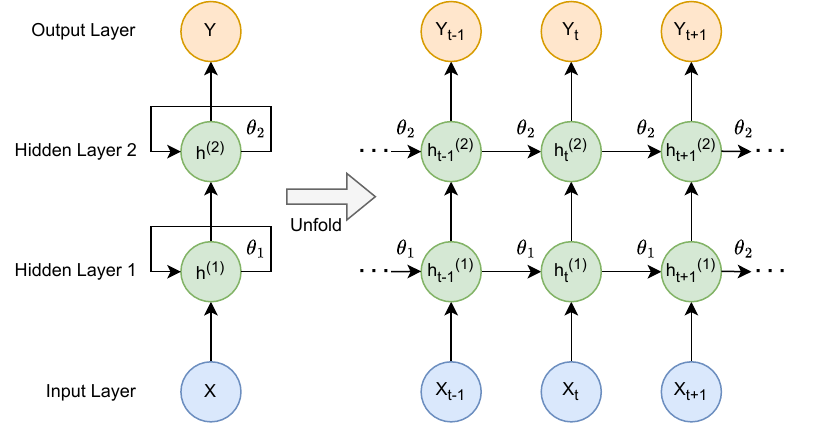}
\caption{A recurrent neural network and its unfolded structure.}
\label{fig_rnn}
\end{figure}

We employ a 2-layer RNN with 64 neurons per layer and Tanh activation functions as the ``foundation model'' architecture. A simple RNN cell follows:
\begin{subequations}
\begin{align}
h_t & = g_1(W^{(x)}x_t + \theta h_{t-1} + b^{(h)}) \\
y_t & = g_2(W^{(y)}h_t + b^{(y)})
\end{align}
\label{eq_rnn}
\end{subequations}
where $h_t$ is the hidden state at time step $t$, $x_t$ is the input at time step $t$, $h_{t-1}$ is the hidden state from the previous time step, $W^{(x)}$, $\theta$ and $W^{(y)}$ are weight matrices for the input, hidden state, and output respectively, $b^{(h)}$ and $b^{(y)}$ are the bias vectors for the hidden state and output respectively, and $y_t$ is the output at time step $t$. The architecture of a 2-layer RNN is illustrated in Fig. \ref{fig_rnn}. During meta-training, the following input-output configurations are utilized based on the type of training sample, summarized as follows: (1) for a CSTR, the inputs consist of  {$[T_t, C_{A_t}, \Delta Q_t, \Delta C_{{A0}_t}]$} at the current time step $t$, and the outputs entail the state trajectory $[T_{t+1, \ldots, n}, C_{A_{t+1, \ldots, n}}]$ over the subsequent $n$ time steps; (2) for a BR, the inputs consist of  {$[T_t, C_{A_t}, \Delta Q_t, \Delta Q_t]$} at the current time step $t$, and the outputs entail the state trajectory $[T_{t+1, \ldots, n}, C_{A_{t+1, \ldots, n}}]$ over the subsequent $n$ time steps; (3) for a PFR, the inputs consist of  {$[T_t, C_{A_t}, \Delta T_{c_t}, \Delta T_{c_t}]$} at the current time step $t$, and the outputs entail the state trajectory $[T_{t+1, \ldots, n}, C_{A_{t+1, \ldots, n}}]$ over the subsequent $n$ time steps.

\begin{figure}[ht]
\centering
\includegraphics[width=\textwidth]{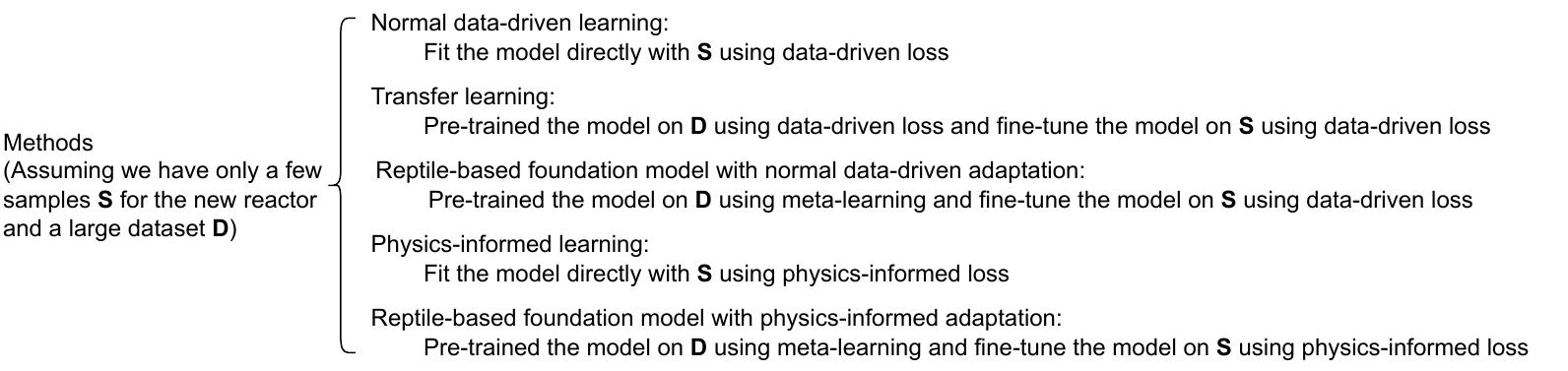}
\caption{Summary of different approaches.}
\label{fig_methods}
\end{figure}

The model is optimized using the Adam optimizer. In the following subsections, we will provide a comprehensive analysis of the few-shot performance with respect to five different approaches on unseen CSTRs, BRs, and PFRs respectively (see Fig. \ref{fig_methods}), formulated as follows:
\begin{itemize}
    \item Normal data-driven learning: This approach involves training a randomly initialized model directly with few-shot samples. It is a standard method where the model learns solely from the provided data without any prior knowledge or pre-training.
    \item Transfer learning: Here, the model is pre-trained on a large dataset containing 500 CSTRs, 500 BRs, and 500 PFRs. This pre-training phase helps the model learn general features and patterns from the data. Then, the pre-trained model is fine-tuned with few-shot samples, allowing it to adapt to specific tasks more efficiently. However, the key difference between transfer learning and meta-learning is that transfer learning focuses on transferring knowledge from one specific task to another related task and develops the pre-trained model following conventional learning algorithms, while meta-learning aims to develop novel learning algorithms such as Reptile that can learn the common pattern across a variety of tasks or domains such that the models can quickly adapt to a new process.
    \item Reptile-based ``foundation model'' with normal data-driven adaptation: In this method, the model is first pre-trained on the 500 CSTRs, 500 BRs, and 500 PFRs dataset using the Reptile learning rule. After pre-training, the ``foundation model'' is fine-tuned with few-shot samples in a normal data-driven manner.
    \item Physics-informed learning: Similar to normal data-driven learning, this approach involves training a randomly initialized model directly with few-shot samples. However, during the training process, a physics-informed loss function is utilized. This loss function incorporates domain-specific knowledge or physical constraints into the learning process, potentially leading to improved generalization and robustness.
    \item Reptile-based ``foundation model'' with physics-informed adaptation: This method combines the benefits of both Reptile-based meta-learning and physics-informed learning. The model is initially pre-trained on the 500 CSTRs, 500 BRs, and 500 PFRs dataset using the Reptile learning rule to enhance its adaptability. Then, during fine-tuning with few-shot samples, a physics-informed loss function is employed to leverage domain-specific knowledge.
\end{itemize}

These approaches vary in their use of prior knowledge, meta-learning techniques, and incorporation of physics-based constraints. By comparing their few-shot performance, we can assess which method is most effective for the given tasks and dataset. The experiments were conducted on Intel Core i7-12700 processor with 64 GB of RAM and a single run for meta-training took about 48~hours, using Python and TensorFlow Keras. Moreover, the source code is available in \url{https://github.com/killingbear999/chemical-reactor-foundation-model}.

\subsection{Few-Shot Adaptation to Unseen CSTRs}
\begin{figure}[ht!]
    \centering
    \begin{subfigure}[t]{0.48\textwidth}
        \centering
        \includegraphics[width=\columnwidth]{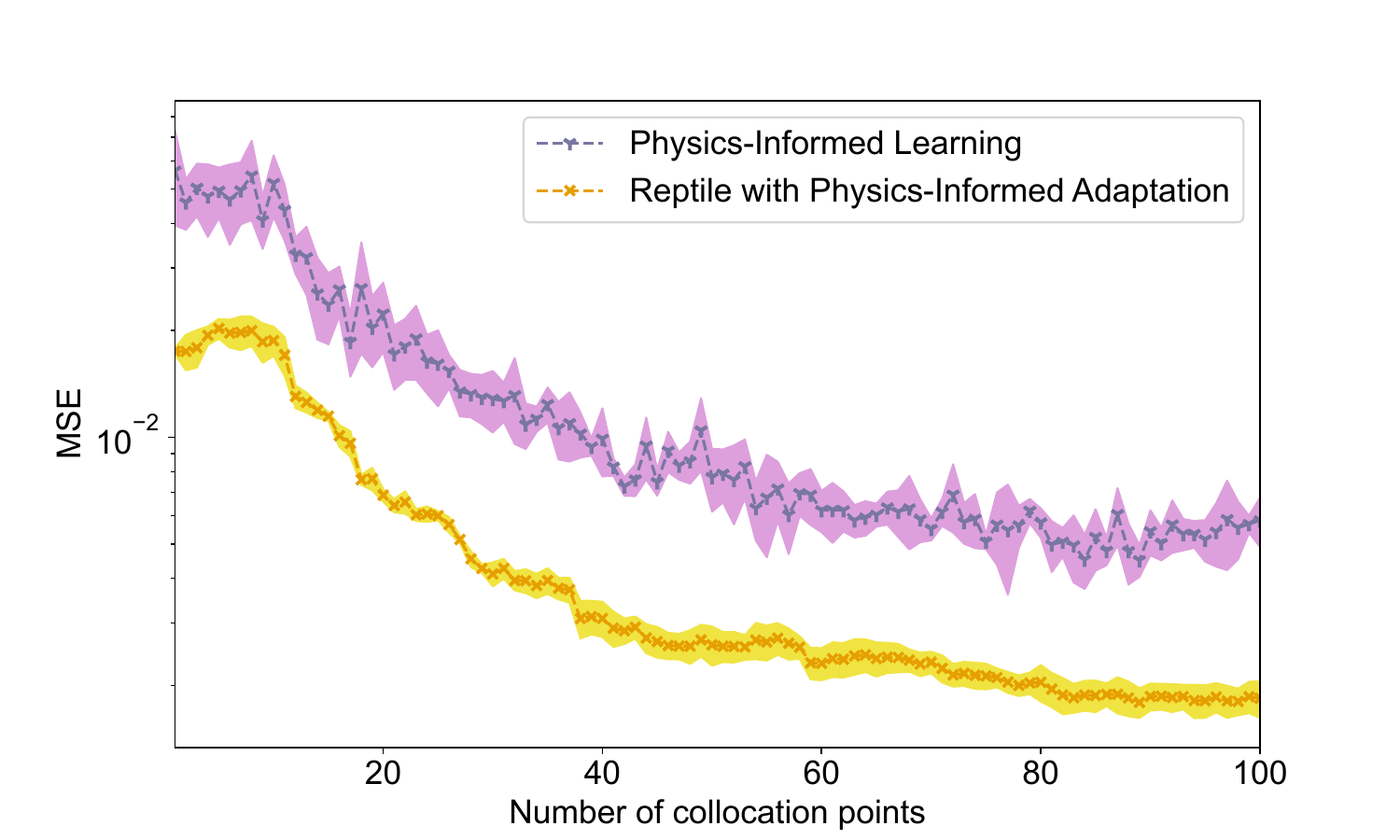}
        \caption{10-shot performance with respect to different number of collocation points on unseen CSTRs.}
        \label{fig_collocation_cstr}
    \end{subfigure}
    ~ 
    \begin{subfigure}[t]{0.48\textwidth}
        \centering
        \includegraphics[width=\columnwidth]{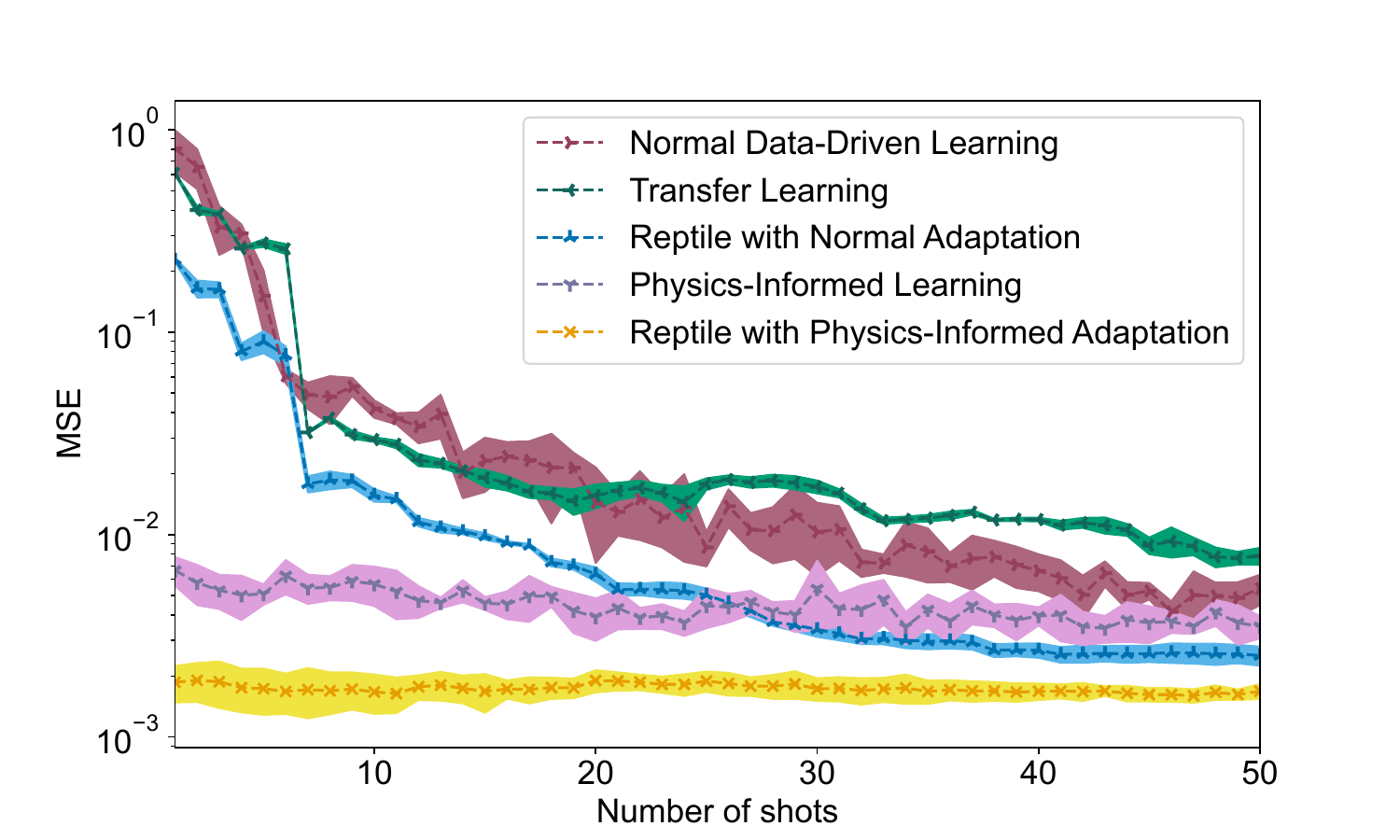}
        \caption{Few-shot performance on unseen CSTRs.}
        \label{fig_fewshots_cstr}
    \end{subfigure}
    \caption{Meta-testing performance on unseen second-order CSTRs.}
    \label{fig_cstr}
\end{figure}

First, we explore how varying the number of collocation points affects model performance in a few-shot setting. As depicted in Fig. \ref{fig_collocation_cstr}, with a fixed number of shots at 10, we observe that physics-informed (PI) approaches reach convergence at around 80 collocation points. Further increasing the number of collocation points does not lead to additional learning. Instead, additional input-output pairs (i.e., increasing the number of shots) are necessary to continue improving the model.

Next, we explore how the number of shots influences model performance across the five aforementioned approaches (i.e., we use 100 collocation points for PI-based methods). From Fig. \ref{fig_fewshots_cstr}, it is evident that the transfer learning approach exhibits the poorest performance. This is because during the pre-training phase, the model encounters difficulties in learning a unified representation due to the significantly different distributions of CSTRs, BRs, and PFRs when using conventional learning algorithms. Conventional physics-informed learning (i.e., without Reptile model) starts with a strong foundation but halts its learning process due to the limited additional information provided by the constrained number of shots, eventually matching the performance of normal data-driven learning with around 50 shots. Reptile with normal adaptation surpasses physics-informed learning after around 30 shots and requires more than 15 shots to achieve satisfactory performance (i.e., reaching an MSE magnitude of $10^{-3}$). In contrast, Reptile with physics-informed adaptation reaches a comparable level with just about 1 shot. Notably, Reptile with physics-informed adaptation maintains the lowest MSE consistently as the number of shots increases.

\subsection{Few-Shot Adaptation to Unseen BRs}
\begin{figure}[ht!]
    \centering
    \begin{subfigure}[t]{0.48\textwidth}
        \centering
        \includegraphics[width=\columnwidth]{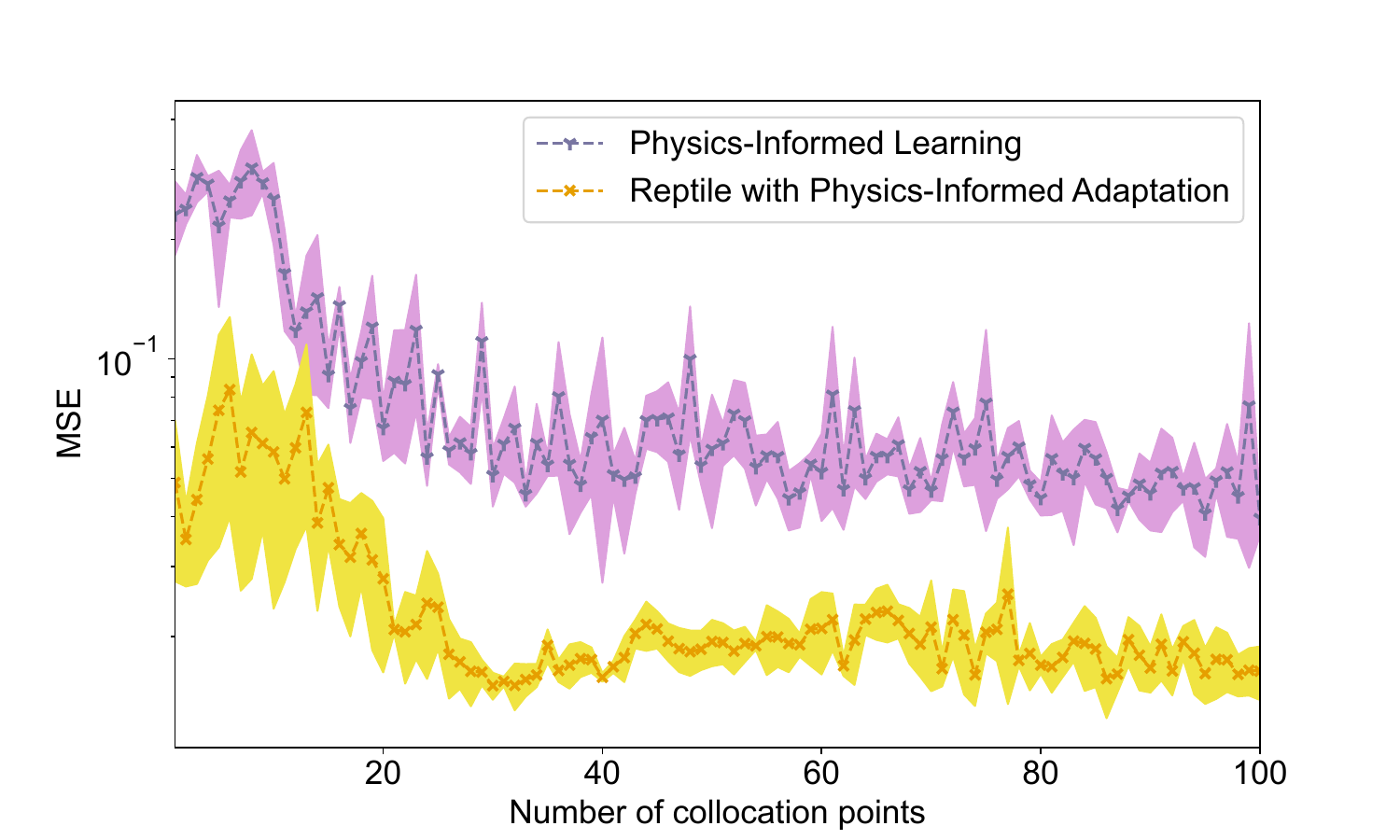}
        \caption{10-shot performance with respect to different number of collocation points on unseen BRs.}
        \label{fig_collocation_batch}
    \end{subfigure}
    ~ 
    \begin{subfigure}[t]{0.48\textwidth}
        \centering
        \includegraphics[width=\columnwidth]{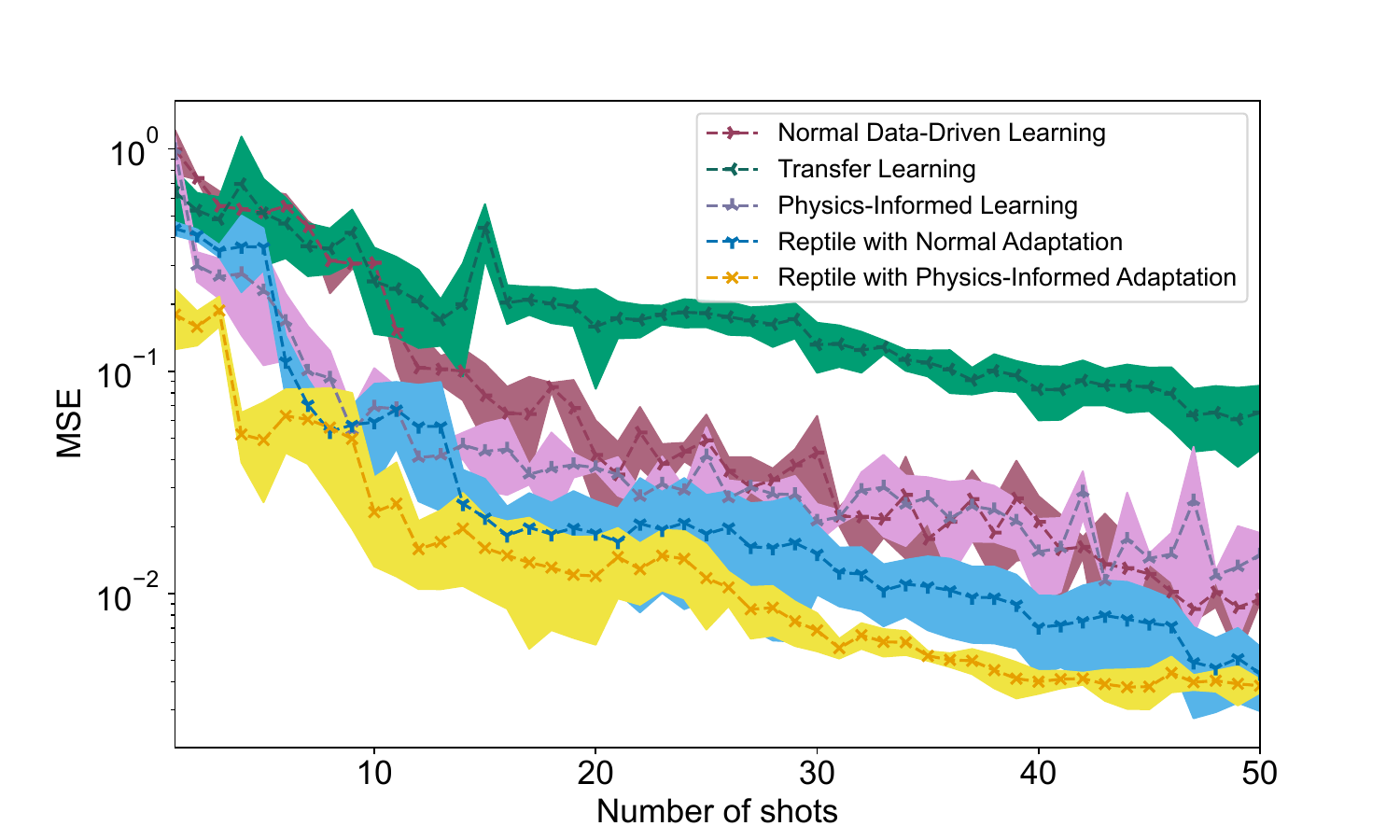}
        \caption{Few-shot performance on unseen BRs.}
        \label{fig_fewshots_batch}
    \end{subfigure}
    \caption{Meta-testing performance on unseen first-order BRs.}
    \label{fig_br}
\end{figure}

First, we investigate how adjusting the number of collocation points influences model performance within a few-shot setting. Similarly to the observations made for CSTRs, illustrated in Fig. \ref{fig_collocation_batch}, when maintaining a fixed number of shots at 10, we find that PI-based approaches converge at approximately 80 collocation points. However, additional increments in the number of collocation points do not yield further learning benefits. Instead, for continued model enhancement, additional input-output pairs (i.e., increasing the number of shots) are required. Furthermore, in contrast to CSTRs, BRs prove to be more challenging to train using PI-based methods. As illustrated in Fig. \ref{fig_collocation_batch}, both PI-based approaches exhibit significant fluctuations, suggesting high sensitivity to the quality of selected collocation points, which aligns with the findings in \cite{rathore2024challenges}.

Next, we investigate how the number of shots influences model performance across the five aforementioned approaches (i.e., using 100 collocation points for PI-based methods). From Fig. \ref{fig_fewshots_batch}, it is apparent that the transfer learning approach performs the poorest, due to difficulties in learning a unified representation during the pre-training phase caused by the distinct distributions of CSTRs, BRs, and PFRs. Conventional physics-informed learning begins with a robust foundation but stalls in its learning trajectory due to the limited additional information from the constrained number of shots, eventually matching normal data-driven learning performance with approximately 25 shots. More importantly, consistent with earlier findings, it is more challenging to train BRs using PI-based methods compared to CSTRs. Reptile with normal adaptation requires approximately 40 shots to achieve satisfactory performance (i.e., reaching an MSE magnitude of $10^{-3}$). Conversely, Reptile with physics-informed adaptation attains a similar level with only about 25 shots. Notably, Reptile with physics-informed adaptation consistently maintains the lowest MSE as the number of shots increases.

\subsection{Few-Shot Adaptation to Unseen PFRs}
\begin{figure}[ht!]
    \centering
    \begin{subfigure}[t]{0.48\textwidth}
        \centering
        \includegraphics[width=\columnwidth]{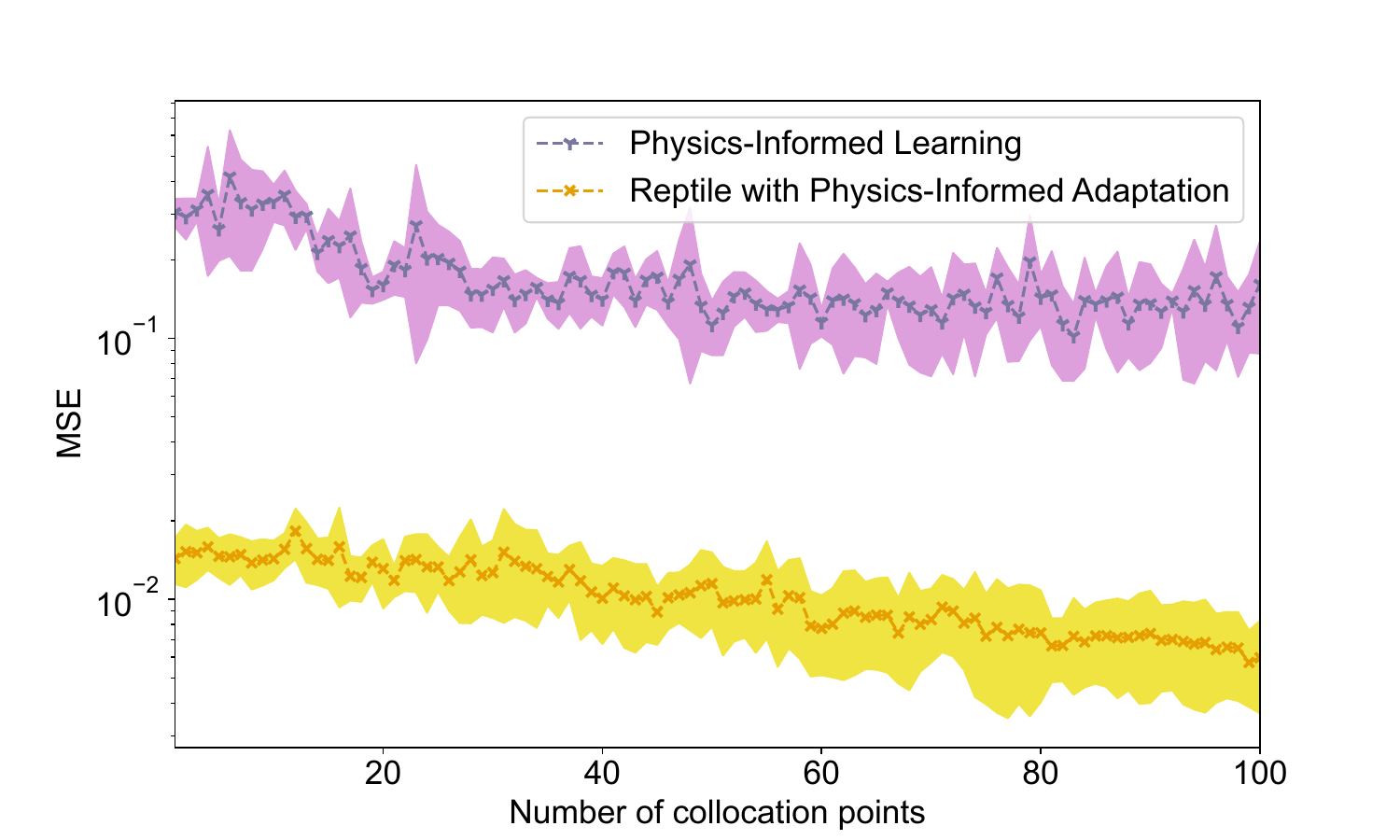}
        \caption{10-shot performance with respect to different number of collocation points on unseen PFRs.}
        \label{fig_collocation_pfr}
    \end{subfigure}
    ~ 
    \begin{subfigure}[t]{0.48\textwidth}
        \centering
        \includegraphics[width=\columnwidth]{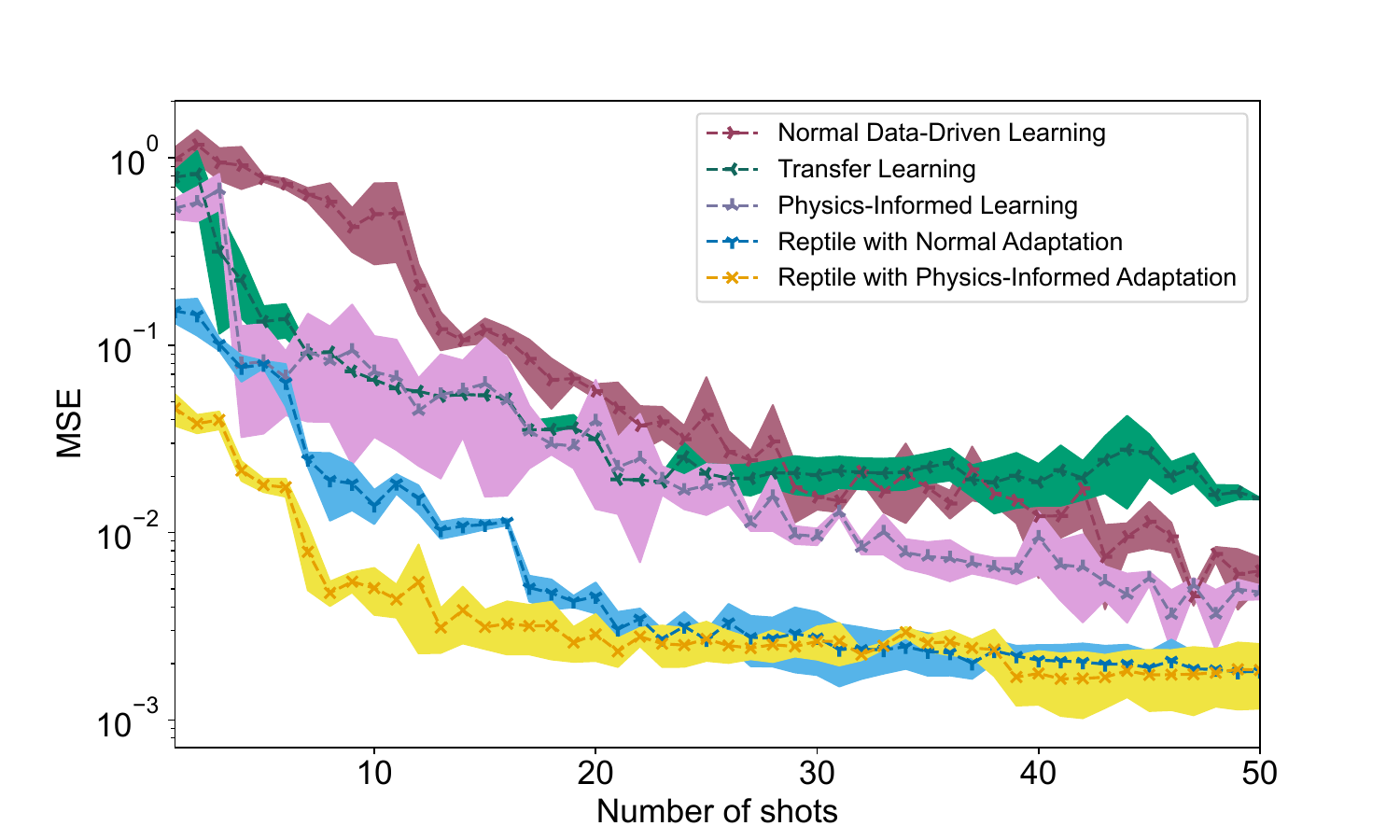}
        \caption{Few-shot performance on unseen PFRs.}
        \label{fig_fewshots_pfr}
    \end{subfigure}
    \caption{Meta-testing performance on unseen first-order PFRs.}
    \label{fig_pfr}
\end{figure}

First, we explore how adjusting the number of collocation points influences model performance within a few-shot setting. As depicted in Fig. \ref{fig_collocation_pfr}, when maintaining a fixed number of shots at 10, we observe that physics-informed learning converges at approximately 50 collocation points, and Reptile with physics-informed adaptation shows little improvement at around 100 collocation points. However, further increases in the number of collocation points do not lead to significant learning benefits. Instead, continued model enhancement requires additional input-output pairs (i.e., increasing the number of shots).
Moreover, unlike CSTRs, PFRs present greater challenges when trained using PI-based methods. As illustrated in Fig. \ref{fig_collocation_pfr}, both PI-based approaches exhibit significant fluctuations, indicating a high sensitivity to the quality of selected collocation points.

Next, we examine how the number of shots influences model performance across the five aforementioned approaches (i.e., with 100 collocation points for PI-based methods). From Fig. \ref{fig_fewshots_pfr}, it is evident that, similar to the previous two cases, the transfer learning approach performs the poorest (i.e., as indicated by its consistently high MSE, particularly when the number of shots exceeds 40). Additionally, conventional physics-informed learning stalls in its learning trajectory due to the limited additional information from the constrained number of shots, and eventually matches normal data-driven learning performance with approximately 50 shots. Furthermore, in line with earlier findings, training PFRs using PI-based methods proves more challenging compared to CSTRs. However, Reptile with normal adaptation requires approximately 17 shots to achieve satisfactory performance (i.e., reaching an MSE magnitude of $10^{-3}$), whereas Reptile with physics-informed adaptation reaches a comparable level with only about 7 shots. Notably, Reptile with physics-informed adaptation consistently maintains the lowest MSE as the number of shots increases.

\subsection{Exemplary trajectory plots}
\begin{figure}[ht!]
    \centering
    \begin{subfigure}[t]{0.48\textwidth}
        \centering
        \includegraphics[width=\columnwidth]{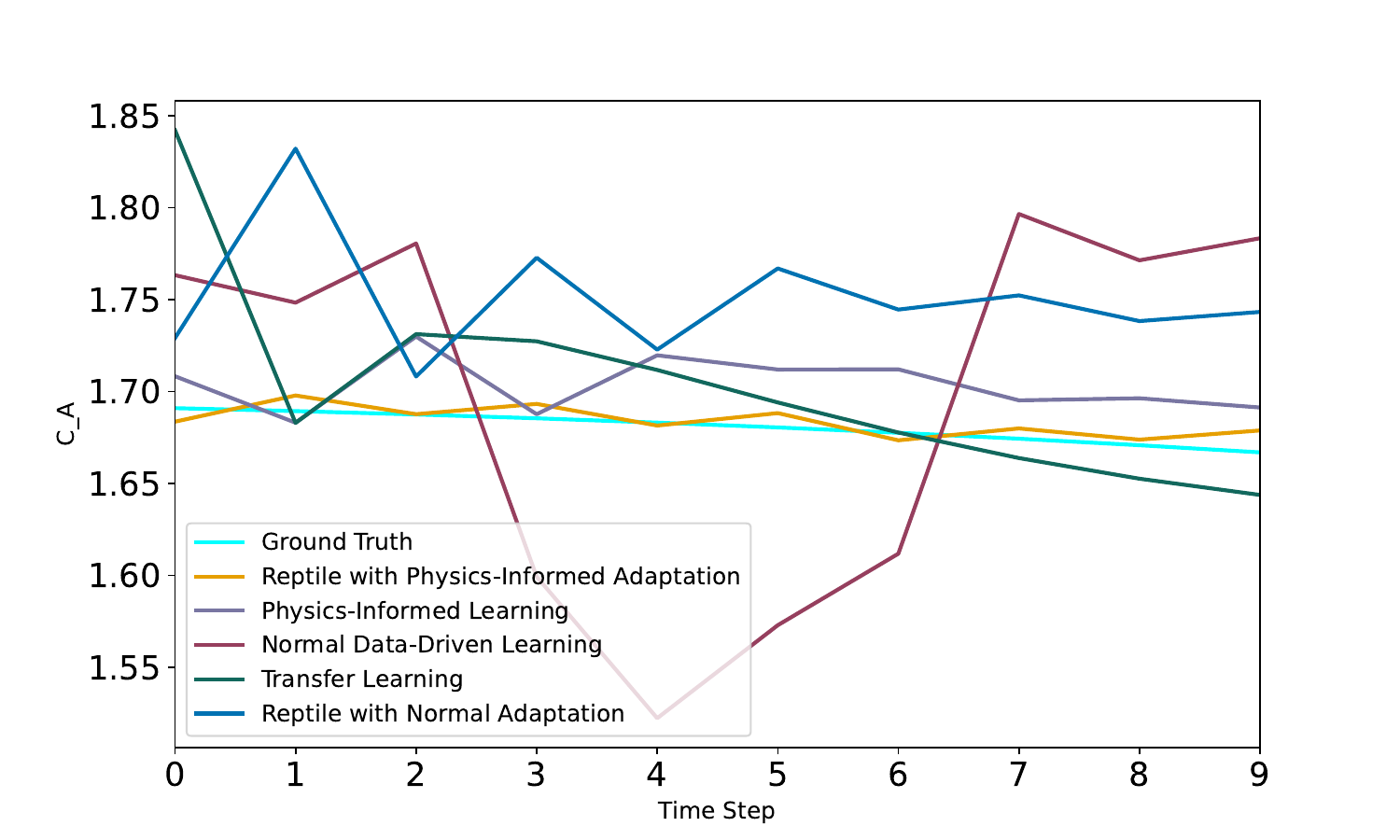}
        \caption{10-shot concentration trajectory plots of an unseen CSTR given a random initial state.}
        \label{fig_cstr_10shots_CA}
    \end{subfigure}
    ~ 
    \begin{subfigure}[t]{0.48\textwidth}
        \centering
        \includegraphics[width=\columnwidth]{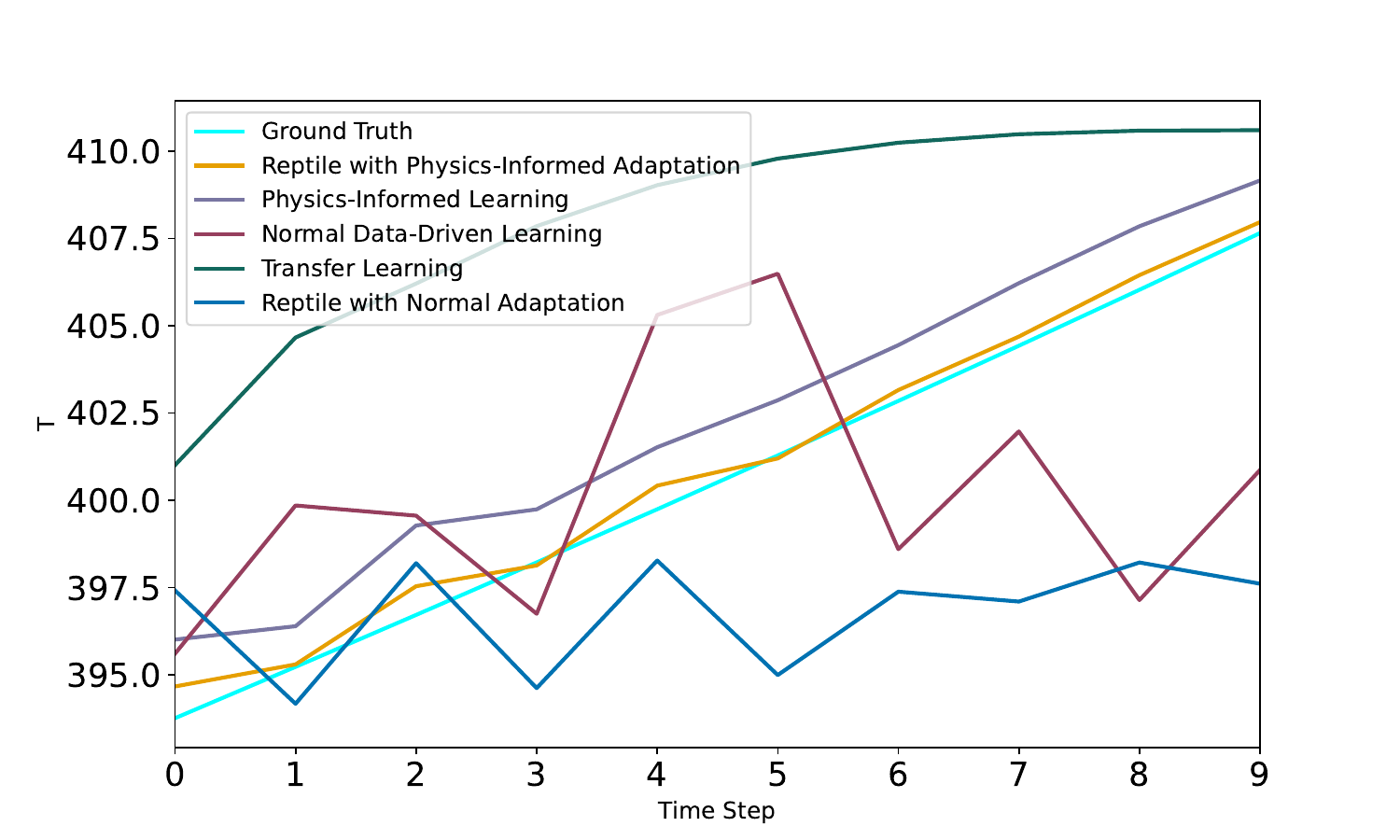}
        \caption{10-shot temperature trajectory plots of an unseen CSTR given a random initial state.}
        \label{fig_cstr_10shots_T}
    \end{subfigure}
    ~
    \begin{subfigure}[t]{0.48\textwidth}
        \centering
        \includegraphics[width=\columnwidth]{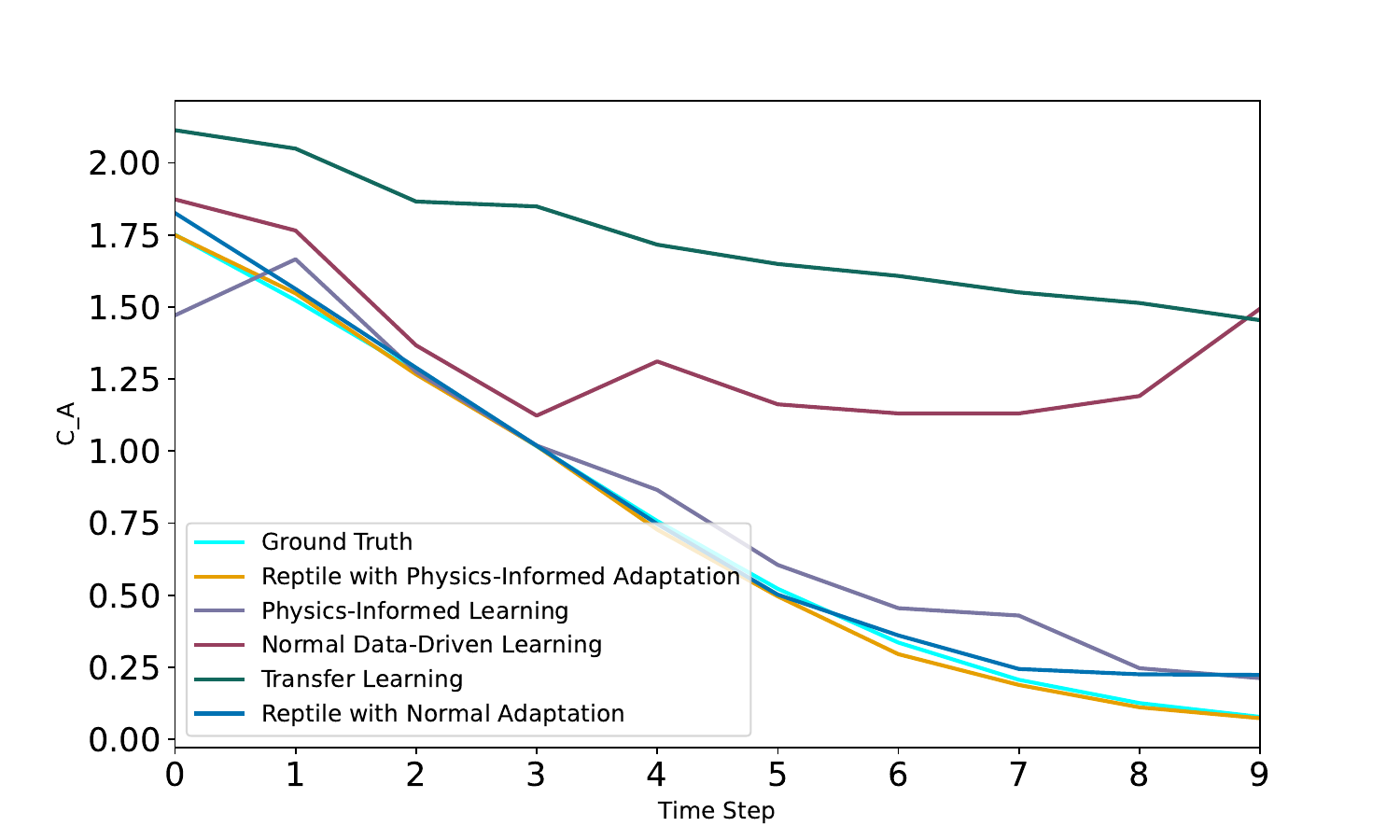}
        \caption{10-shot concentration trajectory plots of an unseen BR given a random initial state.}
        \label{fig_br_10shots_CA}
    \end{subfigure}
    ~
    \begin{subfigure}[t]{0.48\textwidth}
        \centering
        \includegraphics[width=\columnwidth]{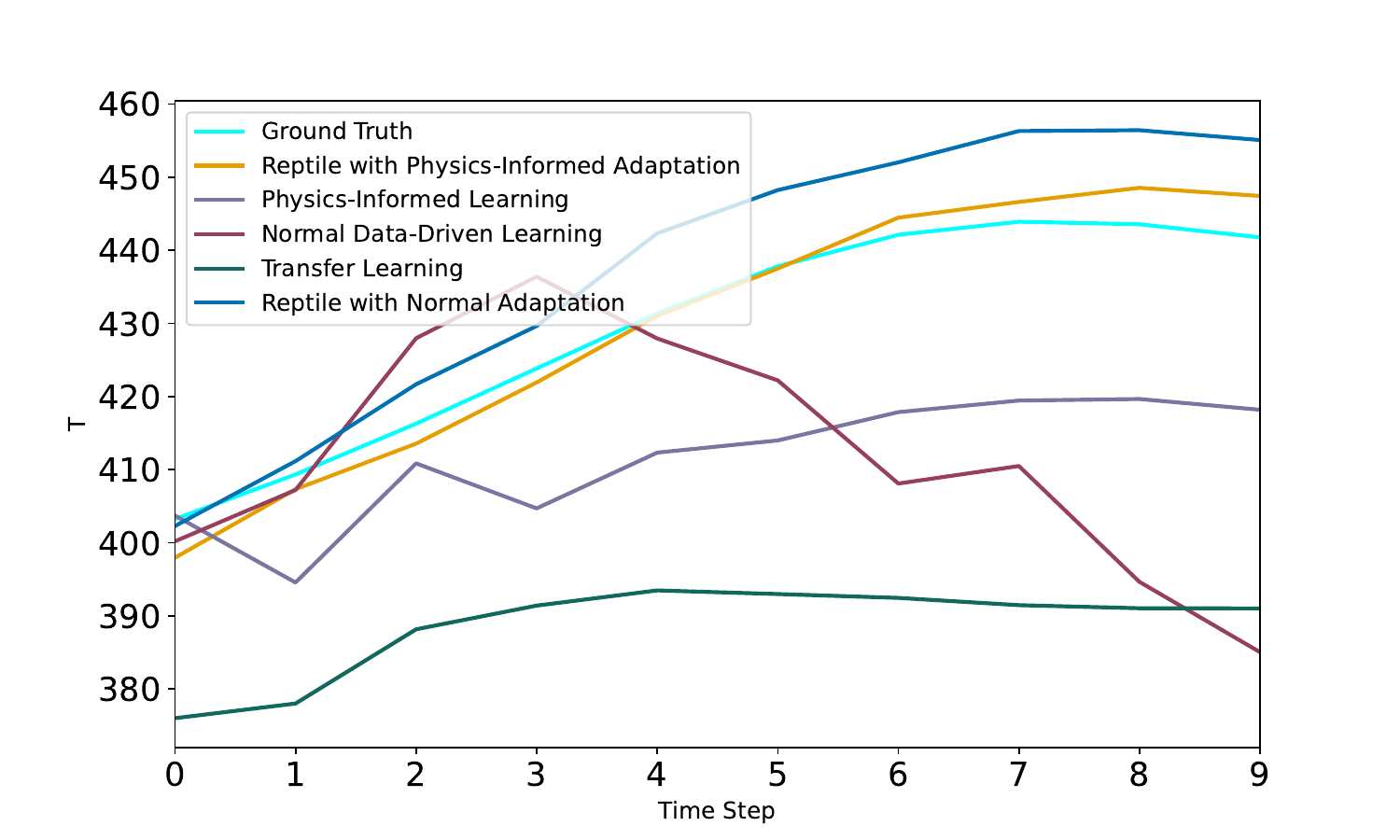}
        \caption{10-shot temperature trajectory plots of an unseen BR given a random initial state.}
        \label{fig_br_10shots_T}
    \end{subfigure}
    ~
    \begin{subfigure}[t]{0.48\textwidth}
        \centering
        \includegraphics[width=\columnwidth]{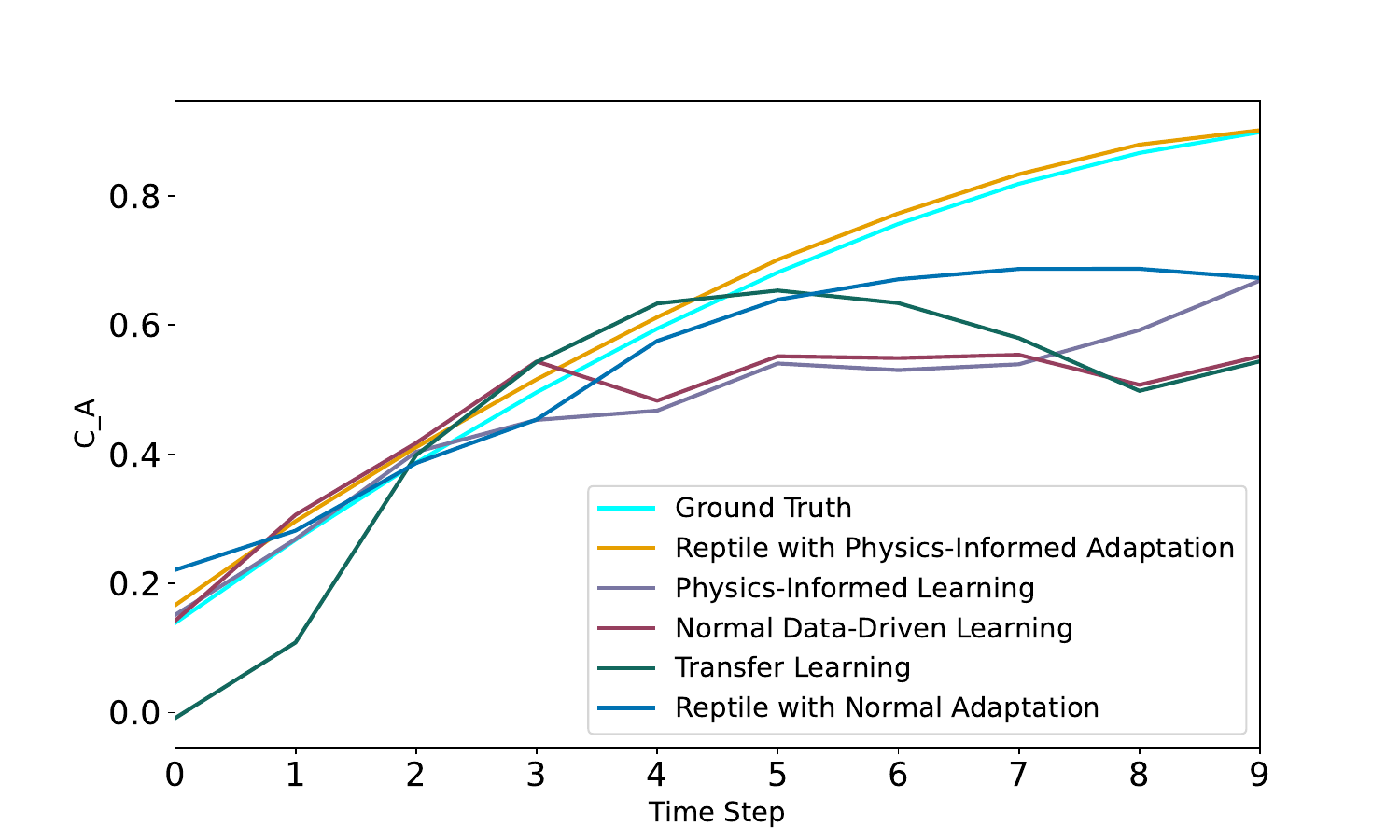}
        \caption{10-shot concentration trajectory plots of an unseen PFR given a random initial state.}
        \label{fig_pfr_10shots_CA}
    \end{subfigure}
    ~
    \begin{subfigure}[t]{0.48\textwidth}
        \centering
        \includegraphics[width=\columnwidth]{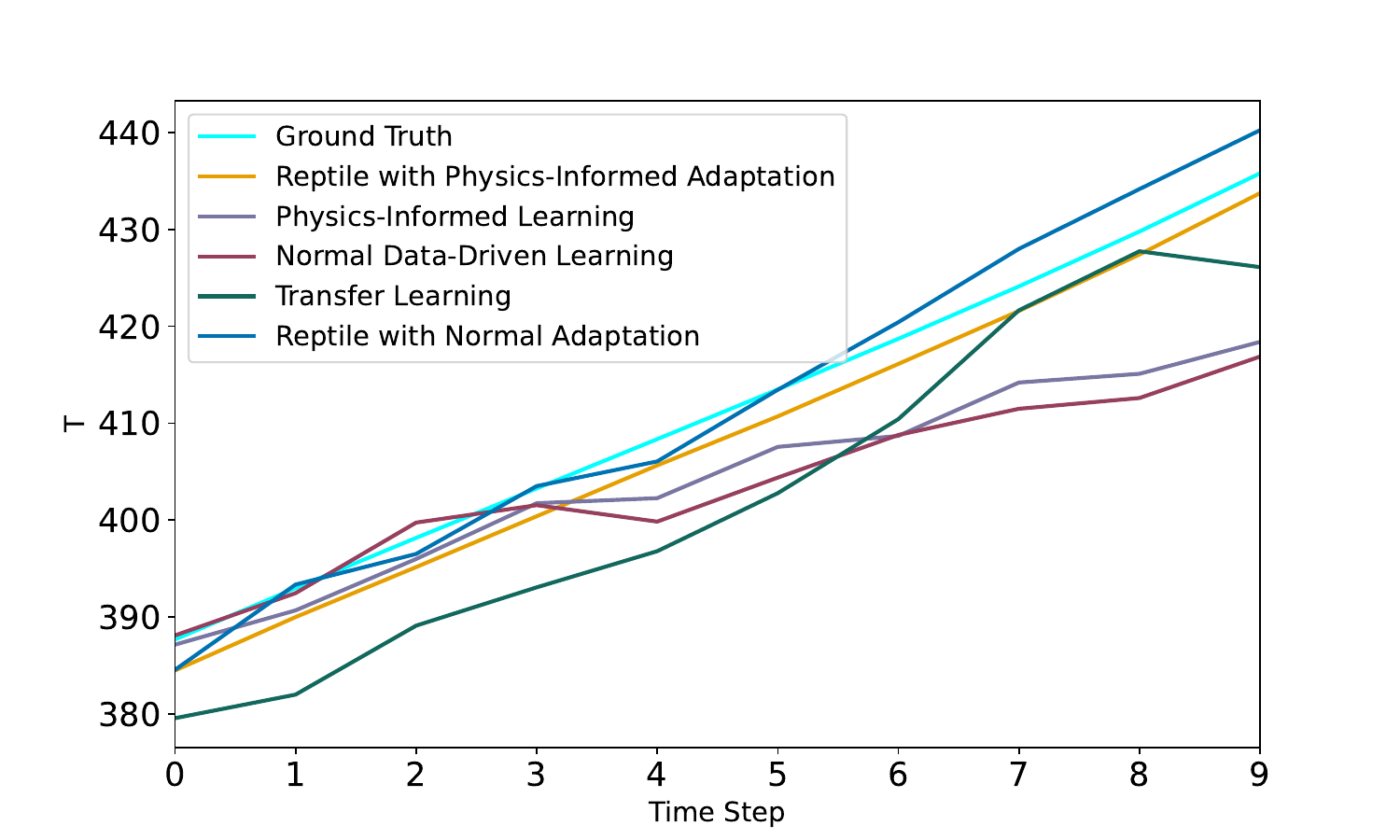}
        \caption{10-shot temperature trajectory plots of an unseen PFR given a random initial state.}
        \label{fig_pfr_10shots_T}
    \end{subfigure}
    \caption{Exemplary trajectory plots after 10-shot adaptation to an unseen CSTR, BR, and PFR.}
    \label{fig_performance}
\end{figure}

To further validate the practical significance of the MSE results, we have included exemplary trajectory plots following a 10-shot adaptation to an unseen CSTR, BR, and PFR, as illustrated in Fig. \ref{fig_performance}. These trajectory plots align with the MSE results, demonstrating that our Reptile approach with physics-informed adaptation achieves the fastest adaptation to the unseen reactor.

\begin{remark}
\label{remark_favor}
Comparing Fig. \ref{fig_fewshots_cstr}, Fig. \ref{fig_fewshots_batch}, and Fig. \ref{fig_fewshots_pfr}, we can observe that the model adapts more efficiently to certain reactor types over others during training and fine-tuning. Specifically, our observations indicate that the ``foundation model'' adapts more quickly to CSTRs, followed by BRs, and then PFRs, suggesting that it learns representations of CSTRs more effectively. As a result, during the adaptation phase, CSTRs achieve the best performance, followed by PFRs, and then BRs. The performance of the different methods is summarized as follows: Reptile with physics-informed adaptation $>$ Reptile with normal adaptation $>$ Physics-informed learning $>$ Normal data-driven learning $>$ Transfer learning.
\end{remark}

\section{Extension to Varying Integer Orders of Reactions}
\label{sec_orders}
\begin{figure}[ht]
\centering
\includegraphics[width=0.6\textwidth]{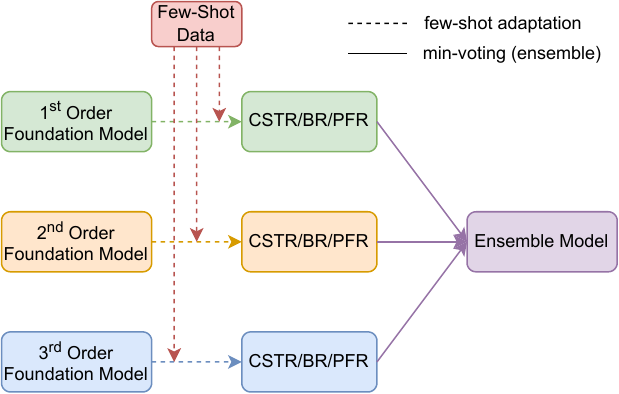}
\caption{High-level system architecture of ensemble model for varying integer orders reactions.}
\label{fig_system_orders}
\end{figure}

Next, we extend the proposed method to handle varying integer reaction orders. Specifically, we meta-train ``foundation models'' for different integer orders of reactions. Consequently, for an unseen integer order of reaction, we integrate their capabilities through ensemble learning, referred to as the ensemble model, with just a few data samples, without requiring precise knowledge of the reaction order, the reaction type, or the complete material and energy balance equations. The high-level system architecture of the ensemble model for varying integer reaction orders is illustrated in Fig. \ref{fig_system_orders}. It can be described in three steps: (1) Train $N$ separate ``foundation model''s, each corresponding to a different integer reaction order; (2) For a new reaction with an integer reaction order, perform few-shot adaptation on all $N$ ``foundation models'' to obtain $N$ intermediate models; (3) Apply min-voting (i.e., select the model with the lowest validation/testing MSE) to determine the final ensemble model, utilizing ensemble learning. 

To validate the few-shot performance of the ensemble model, we compare it against four different approaches on unseen CSTRs, BRs, and PFRs with random integer reaction orders. The experiment setup is consistent with Section \ref{sec_results} with the four approaches outlined as follows:
\begin{enumerate}
    \item Incorrect order Reptile with normal adaptation: In this approach, after obtaining the ``foundation model'' via Reptile, an incorrect order of the ``foundation model'' (i.e., the reaction order used during meta-training differs from that of the unseen meta-testing data) is applied during the meta-testing phase using a standard data-driven adaptation method.
    \item Incorrect order Reptile with physics-informed adaptation: Similar to the previous approach, an incorrect order of the ``foundation model'' is used during the meta-testing phase, but this time adaptation is performed using a physics-informed loss function.
    \item Ensemble Reptile with normal adaptation: This approach follows the ensemble method described in Fig. \ref{fig_system_orders}, using a standard data-driven adaptation for the final ensemble model.
    \item Ensemble Reptile with physics-informed adaptation: Similar to the ensemble Reptile with normal adaptation, but a physics-informed loss function is applied during the adaptation phase for the final ensemble model.
\end{enumerate}

\begin{figure}[ht!]
    \centering
    \begin{subfigure}[t]{0.48\textwidth}
        \centering
        \includegraphics[width=\columnwidth]{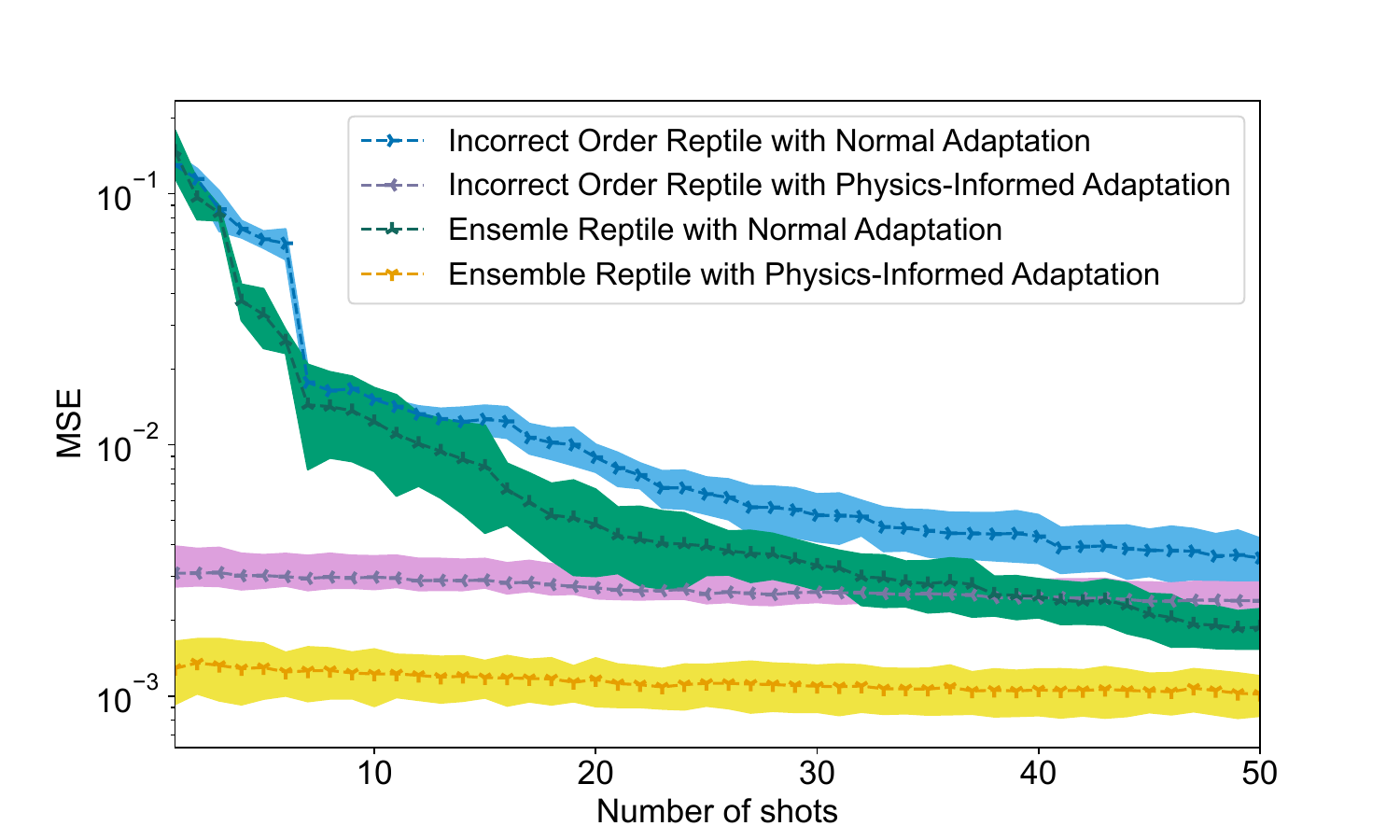}
        \caption{Few-shot performance on unseen CSTRs of random integer orders reactions.}
        \label{fig_cstr_orders}
    \end{subfigure}
    ~ 
    \begin{subfigure}[t]{0.48\textwidth}
        \centering
        \includegraphics[width=\columnwidth]{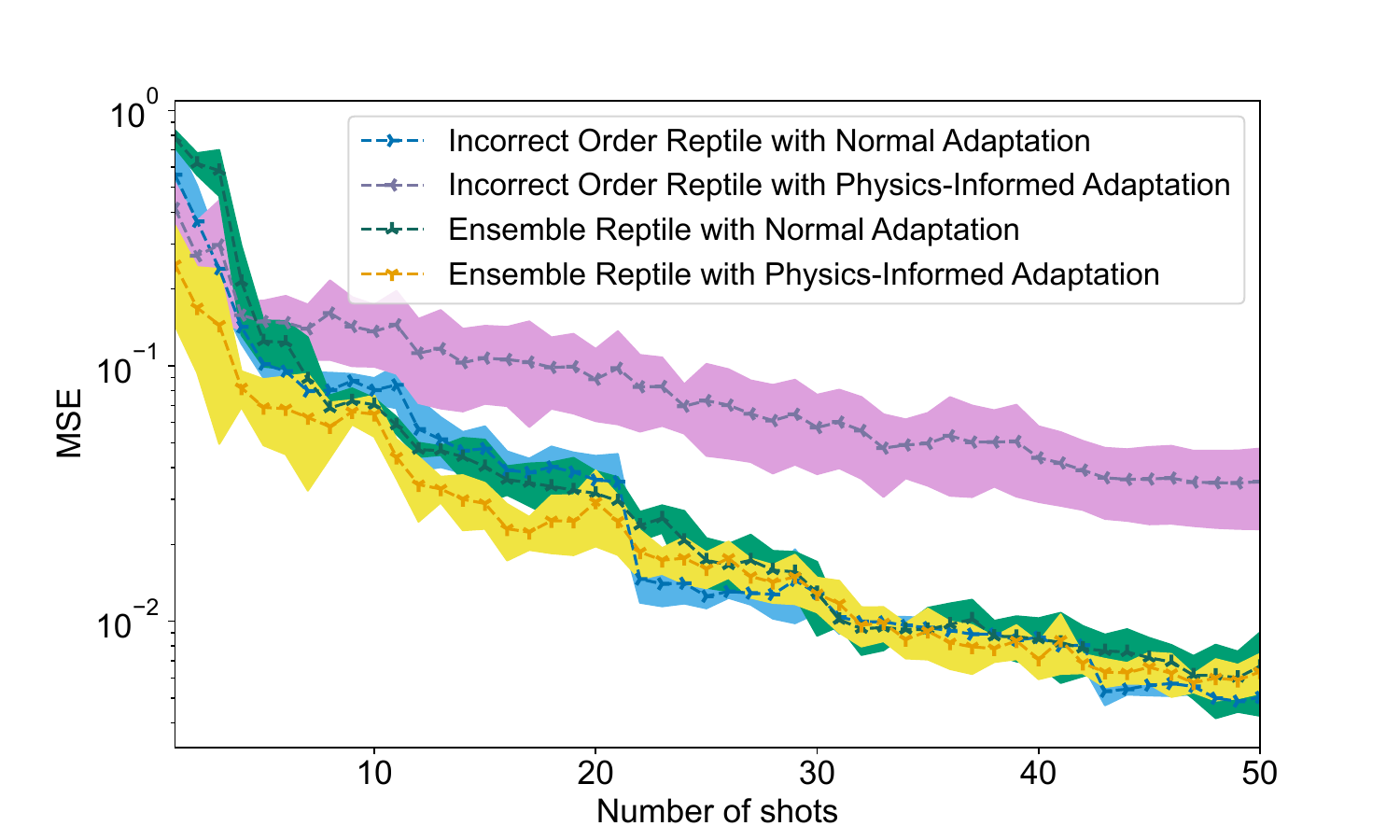}
        \caption{Few-shot performance on unseen BRs of random integer orders reactions.}
        \label{fig_br_orders}
    \end{subfigure}
    ~
    \begin{subfigure}[t]{0.48\textwidth}
        \centering
        \includegraphics[width=\columnwidth]{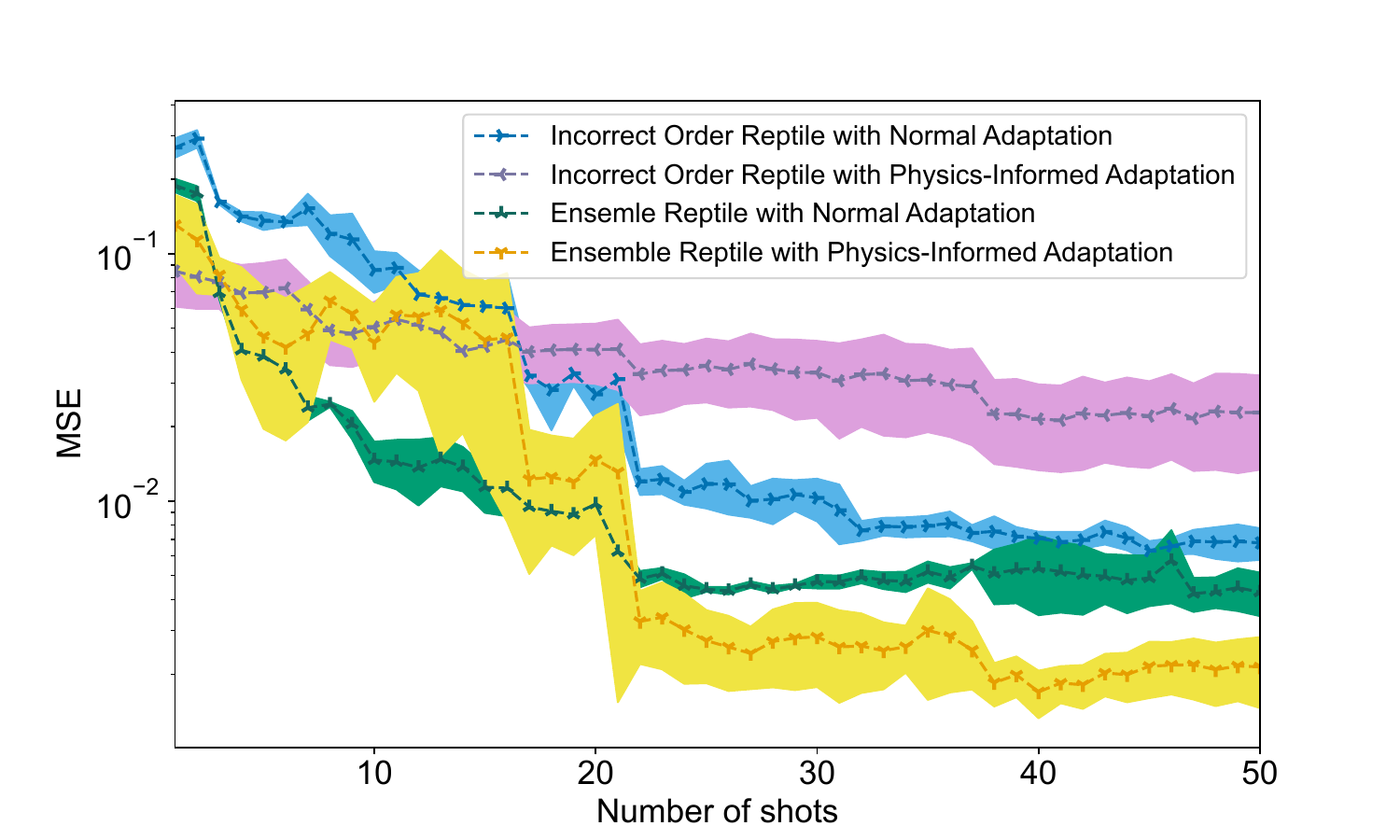}
        \caption{Few-shot performance on unseen PFRs of random integer orders reactions.}
        \label{fig_pfr_orders}
    \end{subfigure}
    \caption{Meta-testing performance on unseen random integer orders CSTRs, BRs, and PFRs.}
    \label{fig_performance_orders}
\end{figure}

The few-shot performance on unseen CSTRs, BRs, and PFRs with random integer reaction orders is presented in Fig. \ref{fig_performance_orders}. As shown in Fig. \ref{fig_cstr_orders} and Fig. \ref{fig_pfr_orders}, the ensemble Reptile with physics-informed adaptation performs the best for unseen CSTRs and PFRs. However, for unseen BRs, the few-shot performance of the ensemble Reptile with physics-informed adaptation is comparable to other methods, such as incorrect order Reptile with normal adaptation and ensemble Reptile with normal adaptation, as shown in Fig. \ref{fig_br_orders}. Moreover, during the adaptation phase, we observe that CSTRs achieve the best performance, followed by PFRs, and then BRs. This aligns with the findings discussed in Section \ref{sec_results} for fixed reaction orders, as noted in Remark \ref{remark_favor}.

Notably, for CSTRs, the ensemble Reptile with normal adaptation performs worse than incorrect-order Reptile with physics-informed adaptation. This is likely because the ``foundation model'' is better suited to CSTRs than to PFRs and BRs, as observed in Fig. \ref{fig_fewshots_cstr}, Fig. \ref{fig_fewshots_batch}, and Fig. \ref{fig_fewshots_pfr}. Thus, even with an incorrect reaction order in the ``foundation model'', it can still quickly adapt to unseen CSTR reactions. In summary, for a well-trained ``foundation model'' (particularly with respect to CSTRs), incorporating a physics-informed loss term accelerates adaptation. However, for a ``foundation model'' that is not as well trained (e.g., with respect to BRs and PFRs), the physics-informed loss term can potentially decelerate adaptation, leading to increased confusion for the model. 

\begin{remark}
\label{remark_collocation_points}
Currently, collocation points are selected randomly. For a fixed number of shots (e.g., 10 shots for adaptation), the choice of collocation points can significantly impact adaptation performance. Suppose that we select 100 collocation points and have two distinct sets of real data samples: Set A and Set B, each containing 10 different samples. These 100 collocation points may perform better with Set A than with Set B, suggesting an underlying correlation between the collocation points and the real data samples used. This indicates that the selection of collocation points is crucial to adaptation performance, especially for BRs and PFRs. As seen in Fig. \ref{fig_br_orders} and Fig. \ref{fig_pfr_orders}, even though the curves for physics-informed approaches show a decreasing trend, they sometimes fluctuate, implying that the chosen collocation points may not align well with the selected shots. To address this issue, a potential improvement is to monitor the adaptation training loss and resample the collocation points if the loss remains high or fails to decrease in subsequent epochs.
\end{remark}

\begin{remark}
To further extend the proposed ``foundation model'' to incorporate varying  {fractional} reaction orders, the meta-training dataset can be expanded to include random fractional reaction orders. During the meta-testing phase, physics-informed adaptation is then applied to facilitate accurate learning and adaptation.
\end{remark}

\section{Limitations and Future Works}
\label{sec_limitations}
Our study demonstrates the model's ability to adapt to unseen reaction kinetics within the studied reactor types - CSTRs, BRs, and PFRs. However, we \textbf{do not claim universal applicability to all chemical processes} in this work. Achieving full generalizability would require pretraining on a significantly broader set of reaction systems and reactor configurations. This work serves as an initial framework toward that goal, with future research aimed at expanding the training dataset to encompass a more diverse range of chemical processes.

The efficacy of our proposed method is heavily influenced by the challenges associated with training PI-based methods for specific tasks due to the use of estimated parameter values in the physics-informed loss terms. Notably, BRs and PFRs pose greater difficulty for PI-based training, resulting in less discernible differences between our Reptile-based ``foundation model'' with physics-informed adaptation and the one with normal data-driven adaptation, compared to CSTRs, where the distinction is more pronounced. To address this, inspired by numerical integration, we propose integrating the driving equations directly into the recurrent cells as a different way to incorporate physical prior into the model, rather than incorporating them into the loss term as a soft constraint. Another challenge with PI-based methods, as discussed in Remark \ref{remark_collocation_points}, lies in the selection of collocation points. Furthermore, our focus on the first discretization element on PFR was primarily driven by ease of validation and computational efficiency. While analyzing behavior further along the reactor could yield additional insights and enrich the dataset, it is beyond the scope of this study (the focus of this work is not to explore methods for data generation or simulate various reactor configurations under different conditions). 

Moreover, the current reactor model is idealized and assumes single-phase homogeneous mixing with a simple $A \rightarrow B$ reaction. Several limitations should be noted. The model does not account for multi-step reactions, parallel pathways, or recycling streams, which are common in industrial settings. The hydrodynamics are also simplified, as PFRs are modeled with unidirectional flow, without considering backflow or axial dispersion. Furthermore, the reaction kinetics are assumed to be intrinsic and ideal, without accounting for mass transfer limitations. Another key limitation is the lack of explicit temperature dependency, meaning that variations in density, enthalpy, and reaction kinetics due to temperature fluctuations are not directly modeled. Additionally, this study is simulation-based due to the lack of publicly available industrial datasets. Unlike fields such as computer vision or natural language processing, where large-scale datasets are widely shared, chemical engineering data is often proprietary, making large-scale data collection and model training more challenging. 

\begin{figure}[ht]
\centering
\includegraphics[width=0.8\textwidth]{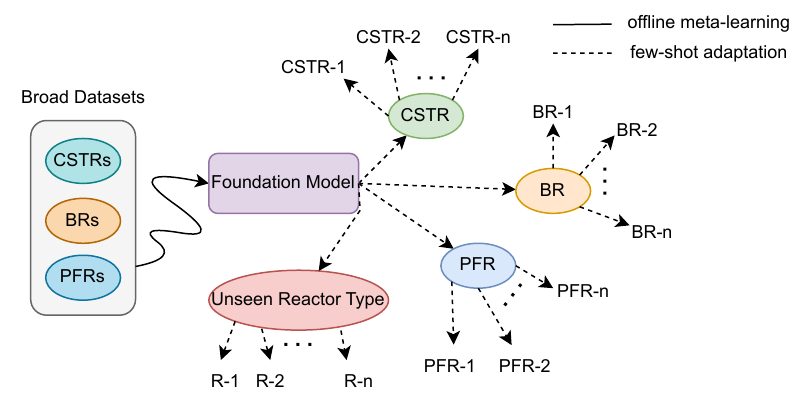}
\caption{High-level system architecture of future ``foundation model'' for various chemical reactors.}
\label{fig_future}
\end{figure}

Future work will focus on improving the model's accuracy and applicability by addressing these limitations. Expanding to more complex reactor systems (see Fig. \ref{fig_future}), such as catalytic fixed-bed reactors and fluidized bed reactors, is an important direction, particularly in systems with multiphase flow, catalyst deactivation, and strong transport-reaction coupling. Incorporating hybrid modeling approaches and physics-based constraints may further enhance generalization to these more intricate reactor types. Additionally, extending the model to handle more realistic reaction networks, including parallel and series reactions, competing pathways, and interacting species, will be crucial for broader applicability. Another avenue for improvement is the explicit modeling of temperature-dependent properties by integrating empirical correlations or physics-informed constraints to capture variations in key physical and chemical parameters.

To further improve generalizability, future research will also leverage Aspen HYSYS, an industry-grade process simulator, to generate additional training data from high-fidelity chemical process models. Furthermore, validating the model on real-world experimental data will be critical in assessing its adaptability to previously unseen reactor types with unknown reaction orders. Finally, while minimizing MSE is a key performance metric, real-world applications must account for noise tolerance. Many downstream tasks, such as neural network-based model predictive control, are inherently robust to measurement noise, meaning that small variations in MSE do not necessarily translate to significant performance differences. Future studies will further evaluate the trade-offs between prediction accuracy and practical usability in noisy environments. By addressing these challenges, we aim to enhance the ``foundation model's'' adaptability, making it more applicable to real-world chemical processes beyond the reactor systems studied in this work.

\section{Conclusion}
\label{sec_conclusion}

In this study, we take a step toward developing foundation models for nonlinear chemical process modeling in reactors. By leveraging meta-learning and physics-informed adaptation, we propose a framework that enables rapid adaptation to previously unseen chemical reactions in CSTRs, BRs, and PFRs with minimal training data. Our approach demonstrates improved few-shot adaptation compared to conventional methods, including data-driven learning, physics-informed learning, transfer learning, and standard meta-learning. While this work does not claim to fully establish a foundation model, it highlights a promising direction for improving generalizability in chemical reactor modeling. This framework has the potential to reduce data requirements and enhance adaptability, which could benefit applications such as process optimization and control in chemical engineering.

\section{Acknowledgments}
Financial support from the MOE-Tier 1 Grant (A-8002899-00-00) is gratefully acknowledged.

\newpage
\bibliography{reference}

\end{document}